\documentclass[letterpaper]{article} 
\usepackage{aaai24}  
\usepackage{times}  
\usepackage{helvet}  
\usepackage{courier}  
\usepackage[hyphens]{url}  
\usepackage{graphicx} 
\usepackage{natbib}  
\usepackage{caption} 
\usepackage{algorithm}
\usepackage{algorithmic}

\usepackage{subfig}
\usepackage{multirow}
\usepackage{enumitem}
\usepackage{amsmath,amssymb}
\usepackage{utfsym}
\usepackage{threeparttable}
\usepackage{bm}
\usepackage[switch]{lineno}

\usepackage{newfloat}
\usepackage{listings}
\DeclareCaptionStyle{ruled}{labelfont=normalfont,labelsep=colon,strut=off} 
\floatstyle{ruled}
\newfloat{listing}{tb}{lst}{}
\floatname{listing}{Listing}
\title{LGMRec: Local and Global Graph Learning for Multimodal Recommendation}
\author{Zhiqiang Guo\textsuperscript{\rm 1}, Jianjun Li\textsuperscript{\rm 1}\thanks{Corresponding author.}, Guohui Li\textsuperscript{\rm 2}\footnotemark[1], Chaoyang Wang\textsuperscript{\rm 3}, Si Shi\textsuperscript{\rm 4}, Bin Ruan\textsuperscript{\rm 1}}
\affiliations{
	\textsuperscript{\rm 1} School of Computer Science and Technology, Huazhong University of Science and Technology, Wuhan, China\\
	\textsuperscript{\rm 2} School of Software Engineering, Huazhong University of Science and Technology, Wuhan, China\\
	\textsuperscript{\rm 3} Wuhan Digital Engineering Institute, Wuhan, China\\
	\textsuperscript{\rm 4} Guangdong Laboratory of Artificial Intelligence and Digital Economy (SZ), Shenzhen, China\\
	\{zhiqiangguo, jianjunli, guohuili\}@hust.edu.cn, sunwardtree@outlook.com, shisi@gml.ac.cn, binruan0227@gmail.com \\
}

\begin{document}
	\maketitle
	
	\begin{abstract}
		The multimodal recommendation has gradually become the infrastructure of online media platforms, enabling them to provide personalized service to users through a joint modeling of user historical behaviors (e.g., purchases, clicks) and item various modalities (e.g., visual and textual). The majority of existing studies typically focus on utilizing modal features or modal-related graph structure to learn user local interests. Nevertheless, these approaches encounter two limitations: (1) Shared updates of user ID embeddings result in the consequential coupling between collaboration and multimodal signals; (2) Lack of exploration into robust global user interests to alleviate the sparse interaction problems faced by local interest modeling. To address these issues, we propose a novel \underline{L}ocal and \underline{G}lobal Graph Learning-guided \underline{M}ultimodal \underline{Rec}ommender (LGMRec), which jointly models local and global user interests. Specifically, we present a local graph embedding module to independently learn collaborative-related and modality-related embeddings of users and items with local topological relations. Moreover, a global hypergraph embedding module is designed to capture global user and item embeddings by modeling insightful global dependency relations. The global embeddings acquired within the hypergraph embedding space can then be combined with two decoupled local embeddings to improve the accuracy and robustness of recommendations. Extensive experiments conducted on three benchmark datasets demonstrate the superiority of our LGMRec over various state-of-the-art recommendation baselines, showcasing its effectiveness in modeling both local and global user interests.
	\end{abstract}
	
	\section{Introduction}
	\label{sec:introduction}

With the explosive growth of massive multimedia information (e.g., images, texts, and videos) on online media platforms, such as YouTube and Tiktok, a lot of efforts have been devoted to multimodal recommender systems (MRSs) to assist these platforms in providing personalized services to users. Nowadays, the primary task of MRSs is to design an effective way to integrate item multimodal information into traditional user-item interaction modeling frameworks to capture comprehensive user interests.

Some early studies on MRSs adopt either the linear fusion between item modal features and their ID embeddings~\cite{he2016vbpr,liu2017deepstyle,wei2021contrastive} or the attention mechanism on item modalities~\cite{chen2017attentive,chen2019personalized,liu2019user} to model representations of users and items. However, The efficacy of these models is somewhat constrained as they only model low-order user-item interactions. The surge of research on graph-based recommendations~\cite{wang2019neural,he2020lightgcn,mao2021ultragcn,wu2021self} has sparked a wave of explorations in using graph neural networks (GNN) to enhance multimodal recommendations. These works typically capture higher-order user interests from the user-item graph that integrates multimodal contents~\cite{wei2019mmgcn,wei2020graph,wang2021dualgnn,yi2022multi,tao2022self,wei2023multi}, or construct modality-aware auxiliary graph structures to transfer multimodal knowledge into item and user embeddings~\cite{zhang2021mining,zhang2022latent,zhou2023bootstrap}.

\begin{figure}[t]
	\centering
	\includegraphics[width=0.98\linewidth]{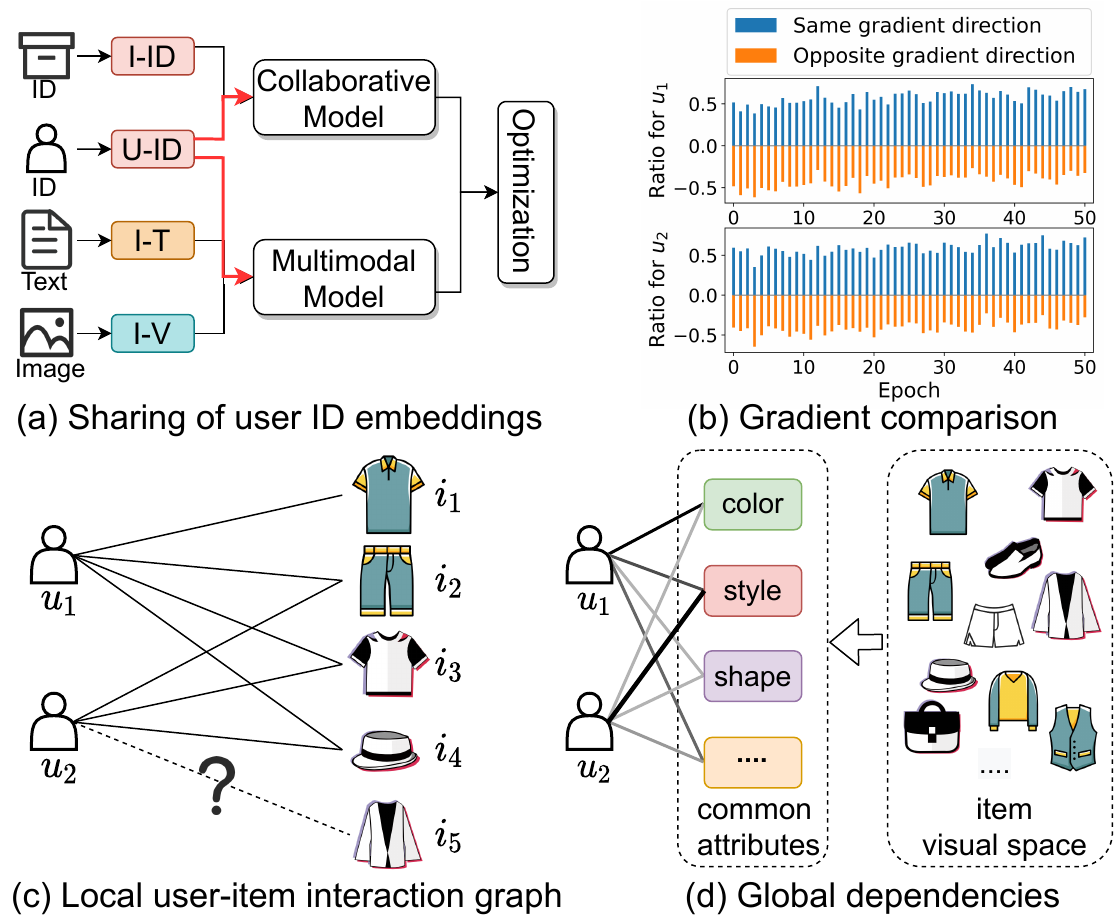}
	\caption{Illustrations of (a) sharing of user ID embeddings, (b) the gradient comparison of user ID embeddings updated from different models during training, (c) local user-item interaction graph, and (d) global dependencies between users and attributions. Darker lines indicate greater user interest.}
	\label{fig:motivation}
\end{figure}

Though achieving remarkable progress, existing studies on MRSs still suffer from the following two limitations in modeling user interests. 
(1) \textit{Coupling}. \textbf{Firstly}, collaboration and multimodal information provide different avenues for exploring user interests. In general, collaborative signals emphasize similar user behavior patterns, while modal knowledge is reflected through content similarity. However, prior works~\cite{wei2019mmgcn,yi2022multi} often overlook this matter and share user ID embeddings in both collaborative and multimodal modeling modules (red line in Figure~\ref{fig:motivation}~(a)) to learn user interests that couple collaborative and multimodal signals. Experimentally, we randomly select two users from the Baby dataset and exhibit the gradient comparison of their ID embeddings (with 64 dimensions) from the collaborative and multimodal modeling modules in Figure~\ref{fig:motivation}~(b). In the early stages of training, the ratio of gradients with opposite directions (orange bar) from the two modules in all dimensions exceeds 50\% for each user, which demonstrates that collaborative and multimodal signals generally have different guidance for user embedding learning\footnote{In fact, approximately 94.26\% of users in Baby dataset present such a situation, that is, more than 50\% of the embedding dimensions have opposite gradient directions during the training process.}. Though this ratio slightly decreases as the training continues, the coupling design still restricts stable updates of user embeddings. 
(2) \textit{Locality}. \textbf{Secondly}, most existing methods~\cite{tao2022self, zhou2023bootstrap} only learn local user interests from the interaction graph (Figure\ref{fig:motivation}~(c)), lacking the exploration of user global interests. Sparse user-item interactions limit their modeling of robust user interests. As shown in Figure\ref{fig:motivation}~(d), user global (general) interests are usually related to item attribute labels that do not rely on the local interactions. Specifically, items usually have multiple common attributions from visual space, such as \textit{color}, \textit{style}, \textit{shape}. Users have different interests in various attributes. For example, $u_1$ may like clothes with bright colors, while $u_2$ prefers a simple style. A method that modeling only local interests may recommend the shirt $i_1$  to $u_2$ based on similar behaviors, i.e., same purchases ($i_2, i_3, i_4$) between $u_1$ and $u_2$. But, the global interests of $u_2$ can provide additional guidance, making it more likely to recommend the outerwear $i_5$ with simple style that match $u_2$'s true interests.

To address the aforementioned issues, we propose a novel \underline{L}ocal and \underline{G}lobal Graph Learning-guided \underline{M}ultimodal \underline{Rec}ommender (LGMRec), which explores capturing and exploiting both local and global representations of users and items to facilitate multimodal recommendation. Specifically, to address the first limitation, we present the local graph embedding module to independently capture collaborative-related and modality-related local user interests by performing message propagation on user-item interaction graphs with ID embeddings and modal features, respectively. In view of the many-to-many dependency relationship between attributes and items is similar to that between hyperedges and nodes in hypergraphs, we further consider each implicit attribute as a hyperedge, and present a global hypergraph embedding module to model hypergraph structure dependencies, so as to address the second limitation. Extensive experimental results on three real-world datasets demonstrate that LGMRec surpasses various recommendation baselines significantly, and verify its effectiveness and robustness in modeling local and global user interests.

	\section{Related Work}
	\label{sec:relatedwork}

\paragraph{Graph-based Recommendation}
The powerful ability of graph neural networks~\cite{kipf2016semi,hu2019hierarchical} in modeling high-order connectivity has greatly promoted the development of recommender systems. Specifically, graph-based recommendation methods model user and item representations by naturally converting the user history interactions into a user-item bipartite graph. Early studies directly inherit the message propagation mechanism of vanilla graph neural network to aggregate high-order neighbor information to represent users and items~\cite{berg2017graph,ying2018graph,wang2019neural}. Later, by simplifying the message propagation process, some graph-based recommendation methods further improve recommendation performance~\cite{chen2020revisiting,he2020lightgcn,mao2021ultragcn}. 
Additionally, some other methods explore more node dependencies to enhance the representations of users and items~\cite{ma2019disentangled,sun2019multi,sun2020neighbor,li2022mdgcf}. Later, contrastive learning is also adopted to enhance graph-based recommendations~\cite{lee2021bootstrapping,wu2021self,yu2022graph,lin2022improving,yang2021enhanced,cai2023lightgcl} to construct contrastive views. However, since no modality features are considered, their modeling abilities are limited by sparse interactions.

\paragraph{Hypergraph learning for Recommendation}
By constructing the hyperedge structure containing more than two nodes, hypergraph learning~\cite{feng2019hypergraph,gao2020hypergraph} can enhance the generalization ability of the model via capturing complex node dependencies. Some recommendation methods~\cite{ji2020dual,wang2020next,he2021click,yu2021self,xia2021self,zhang2022price} try to build hypergraph structures and node-hyperedge connections to capture high-order interaction patterns and achieve substantial performance improvements. 
To further improve performance, several recently developed methods~\cite{xia2022hypergraph,xia2022self} combine self-supervised learning and hypergraph learning to model robust user and item representations. For example, HCCF~\cite{xia2022hypergraph} enhances collaborative filtering with the hypergraph-guided self-supervised learning paradigm. Different from these works that generate hypergraph dependencies via only collaborative embeddings, our work achieves hypergraph structure learning with the modeling of modality-aware global relations.

\begin{figure*}[t]
	\centering
	\includegraphics[width=0.99\linewidth]{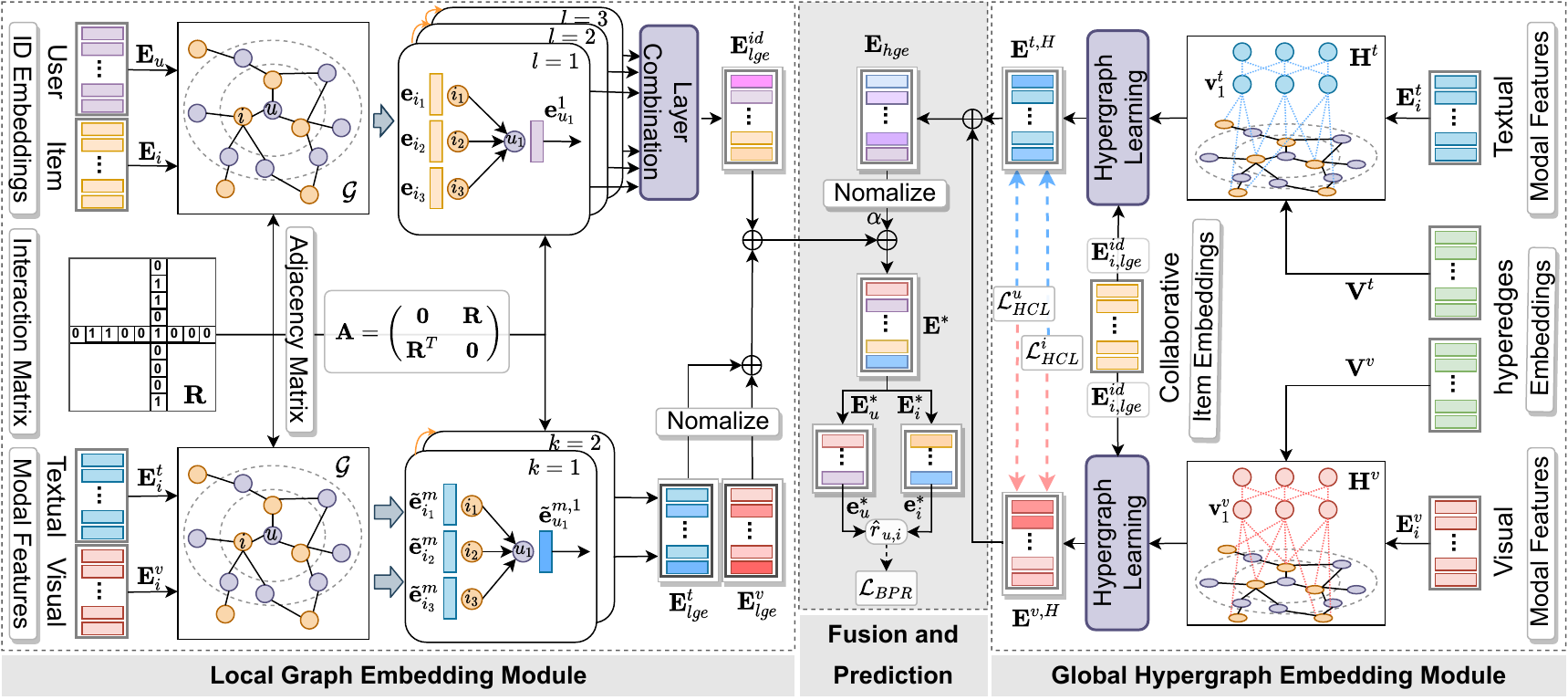}
	\caption{The framework of the proposed LGMRec with visual and textual modalities of items (i.e., $m\in \{v,t\}$).}
	\label{fig:model}
\end{figure*}

\paragraph{Multi-modal Recommendation}
The multi-modal recommendation has become the basic application on online media platforms to provide personalized services to users by analyzing the massive multi-modal information (e.g., images and textual descriptions) and user historical behaviors (e.g., reviews, clicks). Early studies on MRSs usually incorporate multi-modal contents as side information to extend the vanilla CF framework~\cite{he2016vbpr,chen2016context,gao2017unified,du2020learn} or utilize deep autoencoder to model modal features~\cite{guo2022topicvae,liu2022elimrec}. Inspired by the great success of graph-based recommendation methods~\cite{he2020lightgcn,ma2019disentangled,mao2021ultragcn}, some studies directly model user high-order interests on modality-specific interaction graphs~\cite{wei2019mmgcn,wei2020graph,sun2020multi,du2022invariant,kim2022mario}. For instance, MMGCN~\cite{wei2019mmgcn} incorporates modality information into the graph message passing to infer modality-related user preferences. 
Another line utilizes auxiliary semantic graph structures learned from multimodal features to enhance user or item representations~\cite{wang2021dualgnn,zhang2021mining}. For example, LATTICE~\cite{zhang2021mining} is a representative method that exploits modal content similarity to generate auxiliary latent item semantic relations to promote recommendation. Recently, Some works~\cite{wei2021contrastive,yi2022multi,tao2022self,zhang2022latent,zhou2023bootstrap,wei2023multi} introduce contrastive learning into MRSs to model robust user and item representations. 
However, these methods usually perform message passing along the edges of user-item interactions to obtain local user interests, failing to explore modality-aware comprehensive user interests.

	\section{Methodology}
	\label{sec:method}

In this section, we first formulate the problem of multimodal recommendation and present the overall framework of our LGMRec, and then introduce each component in detail.

\subsection{Problem Statement and Overview}
We set the user set as  $\mathcal{U}=\{u\}$ and item set as $\mathcal{I}=\{i\}$. The ID embeddings of each user $u\in \mathcal{U}$ and item $i \in \mathcal{I}$ are denoted as $\mathbf{e}_u \in \mathbb{R}^d$ and $\mathbf{e}_i \in \mathbb{R}^d$, respectively, where $d$ is the embedding dimension. The user-item interactions can be represented as a matrix $\mathbf{R} \in \mathbb{R}^{|\mathcal{U}| \times |\mathcal{I}|}$, in which the element $r_{u,i}=1$ if user $u$ interacts with item $i$, and $r_{u,i}=0$ otherwise. Based on interaction matrix $\mathbf{R}$, we can construct the user-item interaction graph $\mathcal{G}=\{\mathcal{U} \cup \mathcal{I}, \mathcal{E}\}$, where $\mathcal{E}$ is edge set build on observed interactions, i.e., a nonzero $r_{u,i}$ corresponds to an edge between user $u$ and item $i$ on the graph $\mathcal{G}$. Further, we incorporate item multimodal contents and denote the original modality feature of item $i$ generated from pre-trained models as $\mathbf{e}^m_i \in \mathbb{R}^{d_m}$ under modality $m \in \mathcal{M}$, where $\mathcal{M}$ is the set of modalities and $d_m$ denotes the dimension of modal features. In this work, we consider two mainstream modalities, vision $v$ and text $t$, i.e., $\mathcal{M}=\{v,t\}$. Given the above settings, the multimodal recommendation aims to learn a prediction function to forecast the score $\hat{r}_{u,i}$ of an item $i$ adopted by a user $u$ via joint modeling user behaviors and multimodal contents. Formally,
\begin{equation}
 \hat{r}_{u,i} = \textsc{Prediction}\left(\mathbf{R}, \mathbf{E}^{id}, \{\mathbf{E}_i^m\}_{m\in\mathcal{M}}\right)
\end{equation}
where $\textsc{Prediction}(\cdot)$ is the prediction function, $\mathbf{E}^{id}=[\mathbf{e}_{u_1}, \ldots, \mathbf{e}_{u_{|\mathcal{U}|}}, \mathbf{e}_{i_1}, \ldots, \mathbf{e}_{i_{|\mathcal{I}|}}] \in \mathbb{R}^{(|\mathcal{U}|+|\mathcal{I}|) \times d}$ denotes the ID embedding matrix by stacking all the ID embeddings of users and items, $\mathbf{E}_i^m = [\mathbf{e}^m_{i_1}, \ldots, \mathbf{e}^m_{i_{|\mathcal{I}|}}] \in \mathbb{R}^{|\mathcal{I}| \times d_m}$ is the item modal feature matrix under modality $m$.

\textbf{Overview}. As illustrated in Figure~\ref{fig:model}, the framework of LGMRec consists of three major components: (i) Local graph embedding (LGE) module, which adopts GNN to capture collaborative-related and modality-related user local interests on user-item interaction graph with ID embeddings and modal features, respectively; (ii) Global hypergraph embedding (GHE) module, which learns the global user and item representations by capturing the global hypergraph structure dependencies from different item modal feature spaces; and (iii) Fusion and prediction module, which fuses both local and global embeddings to predict final user preference scores for items. 

\subsection{Local Graph Embedding (LGE) Module}
The LGE module is designed to independently learn the collaborative-related and modality-related user and item representations with local topology structure for avoiding unstable updates of user embeddings and promoting decoupled user interest learning. 
\subsubsection{Collaborative Graph Embedding (CGE)}
We first capture the high-order connectivity via the message propagation on the user-item interaction graph with ID embeddings. In particular, the collaborative graph propagation function $\textsc{CGProg}(\cdot)$ in the $(l+1)$-th layer can be formatted as,
\begin{equation}
    \mathbf{E}^{l+1}=\textsc{CGProg}(\mathbf{E}^{l}) = \left(\mathbf{D}^{-\frac{1}{2}} \mathbf{A} \mathbf{D}^{-\frac{1}{2}}\right) \mathbf{E}^{l},
\end{equation} 
where $\textsc{CGProg}(\cdot)$ function inherits the lightweight form of the simplified graph convolutional network~\cite{chen2020revisiting,he2020lightgcn}, $\mathbf{A} \in \mathbb{R}^{(|\mathcal{U}|+|\mathcal{I}|) \times (|\mathcal{U}|+|\mathcal{I}|)}$ is the adjacency matrix constructed from interaction matrix $\mathbf{R}$, and $\mathbf{D}$ is the diagonal matrix of $\mathbf{A}$. Each diagonal element $\mathbf{D}_{j,j}$ in $\mathbf{D}$ denotes the number of nonzero entries in the $j$-th row vector of matrix $\mathbf{A}$. The initial embeddings matrix is set as $\mathbf{E}^{0} = \mathbf{E}^{id}$. Then, we adopt the layer combination~\cite{he2020lightgcn} to integrate all embeddings from hidden layers,
\begin{equation}
	\mathbf{E}^{id}_{lge} =\textsc{Layercomb}\left(\mathbf{E}^0, \mathbf{E}^1, \mathbf{E}^2, \ldots, \mathbf{E}^L\right),
\end{equation} 
where $\mathbf{E}^{id}_{lge} \in \mathbb{R}^{(|\mathcal{U}|+|\mathcal{I}|) \times d}$ is collaborative-related embeddings of users and items with local neighborhood information. We use the mean function to achieve $\textsc{Layercomb}(\cdot)$ for embedding integration.

\subsubsection{Modality Graph Embedding (MGE)}
Considering the semantic differences between modalities, we further independently infer the modality-related embeddings of users and items on the interaction graphs with modal features. The original modal features of items are usually generated from different pre-trained models, e.g., ResNet~\cite{he2016deep}, BERT~\cite{kenton2019bert}, they have different dimensions in different feature spaces. We require the projection of high-dimensional modal feature $\mathbf{e}_i^m$ of each item into a unified embedding space $\mathbb{R}^d$ as,
\begin{equation}
\widetilde{\mathbf{e}}_i^m = \textsc{Transform}(\mathbf{e}_i^m) = \mathbf{e}_i^m \cdot \mathbf{W}_m, \\
\label{equ:projection}
\end{equation}
where $\widetilde{\mathbf{e}}_i^m$ is item $i$'s transformed modal feature, $\textsc{Transform}(\cdot)$ is a projection function parameterized by a transformation matrix $\mathbf{W}_m \in \mathbb{R}^{d_m \times d}$. Due to the difficulty in obtaining user modal information, existing methods often reuse user ID embedding as input for modality-specific graphs, resulting in coupling of collaborative and modal signals. Different from them, we initialize the user modal features by aggregating  item modal features,
\begin{equation}
\widetilde{\mathbf{e}}_u^m = \frac{1}{\left|\mathcal{N}_u\right|}\sum\nolimits_{i \in \mathcal{N}_u} \widetilde{\mathbf{e}}_i^m,
\label{equ:user_modal}
\end{equation}
where $\mathcal{N}_u$ denotes the neighbor set of user $u \in \mathcal{U}$ on user-item interaction graph $\mathcal{G}$. This operation ensures the separate updates of ID embedding and modal features. Thereafter, we can construct the modal feature matrix $\widetilde{\mathbf{E}}^m = [\widetilde{\mathbf{e}}^m_{u_1}, \ldots, \widetilde{\mathbf{e}}^m_{u_{|\mathcal{U}|}}, \widetilde{\mathbf{e}}^m_{i_1}, \ldots, \widetilde{\mathbf{e}}^m_{i_{|\mathcal{I}|}}] \in \mathbb{R}^{(|\mathcal{U}|+|\mathcal{I}|) \times d}$ as initial input $\widetilde{\mathbf{E}}^{m,0}$ to learn modality-related embeddings via implementing a light graph propagation function $\textsc{MGProg}(\cdot)$,
\begin{equation}
\widetilde{\mathbf{E}}^{m,k+1}=\textsc{MGProg}(\widetilde{\mathbf{E}}^{m,k})=\left(\mathbf{D}^{-\frac{1}{2}} \mathbf{A} \mathbf{D}^{-\frac{1}{2}}\right) \widetilde{\mathbf{E}}^{m,k}.
\end{equation} 
Here, we choose high-order modal embeddings $\widetilde{\mathbf{E}}^{m,K}$ in the last $K$-th layer as the modality-related embeddings (i.e., $\mathbf{E}^m_{lge} = \widetilde{\mathbf{E}}^{m,K}$) with local modal information.

\subsection{Global Hypergraph Embedding (GHE) Module} 
The GHE module is designed to capture the modality-aware global representations of users and items against sparse and noisy user behaviors. 
\subsubsection{Hypergraph Dependency Constructing}
Explicit attribute information of item modalities is often unavailable, especially for visual modalities. Hence, we define learnable implicit attribute vectors $\{\mathbf{v}^m_a\}_{a=1}^A$ ($\mathbf{v}^m_a \in \mathbb{R}^{d_m}$) as hyperedge embeddings under modality $m$ to adaptively learn the dependencies between implicit attributes and items/users , where $A$ is the number of hyperedges. Specifically, We obtain hypergraph dependency matrices in low-dimensional embedding space by,
\begin{equation}
\mathbf{H}_i^m = \mathbf{E}_i^m \cdot {\mathbf{V}^m}^{\top}, \quad \mathbf{H}_u^m = \mathbf{A}_u \cdot {\mathbf{H}_i^m}^{\top},
\end{equation}
where $\mathbf{H}_i^m \in \mathbb{R}^{|\mathcal{I}| \times A}$ and $\mathbf{H}_u^m \in \mathbb{R}^{|\mathcal{U}| \times A}$ are the item-hyperedge and user-hyperedge dependency matrices, respectively. $\mathbf{E}_i^m$ is the raw item modal feature matrix, $\mathbf{V}^m = [\mathbf{v}^m_1, \ldots, \mathbf{v}^m_A] \in \mathbb{R}^{A \times d_m}$ is the hyperedge vector matrix, and $\mathbf{A}_{u} \in \mathbb{R}^{|\mathcal{U}| \times |\mathcal{I}|}$ is the user-related adjacency matrix extracted from $\mathbf{A}$. Intuitively, items with similar modal features are more likely to be connected to the same hyperedge. The user-hyperedge dependencies are indirectly derived through the user-item interactions, which implies the user behavior intention, i.e., the more frequently users interact with items under a certain attribute, the more they may prefer the attribute.

To further avoid the negative impact of meaningless relationships, we employ the Gumbel-Softmax reparameterization~\cite{eric2017categorical} to ensure that an item is attached to only one hyperedge as much as possible,
\begin{equation}
	\widetilde{\mathbf{h}}_{i,*}^m = \textsc{Softmax}\left(\frac{\log \bm{\delta}-\log (1-\bm{\delta})+\mathbf{h}_{i,*}^m}{\tau}\right),
\end{equation}
where $\mathbf{h}_{i,*}^m \in \mathbb{R}^A$ is the $i$-th row vector of $\mathbf{H}_i^m$ that reflects the relations between item $i$ and all hyperedges. $\bm{\delta} \in \mathbb{R}^A$ is a noise vector, where each value $\delta_j \sim \text{Uniform}(0,1)$, and $\tau$ is a temperature hyperparameter. Afterwards, we can get the augmented item-attribute hypergraph dependency matrix $\widetilde{\mathbf{H}}_i^m$. By performing similar operations on $\mathbf{H}_u^m$, we can obtain the augmented user-attribute relation matrix $\widetilde{\mathbf{H}}_u^m$.

\subsubsection{\textbf{Hypergraph Message Passing}}
By taking the attribute hyperedge as an intermediate hub, we achieve hypergraph message passing to deliver global information to users and items without being limited by hop distances. Formally,
\begin{equation}
	\mathbf{E}^{m,h+1}_{i} = \textsc{Drop}(\widetilde{\mathbf{H}}_i^m) \cdot \textsc{Drop}(\widetilde{\mathbf{H}}_i^{m \top}) \cdot \mathbf{E}^{m,h}_{i},
\end{equation}
where $\mathbf{E}^{m,h}_i$ is the global embedding matrix of items in the $h$-th hypergraph layer, and $\textsc{Drop}(\cdot)$  denotes a dropout function. We take collaborative embedding matrix $\mathbf{E}^{id}_{i,lge}$ of items as the initial global embedding matrix when $h=0$. Further, we can calculate the global user embedding matrix as,
\begin{equation}
\mathbf{E}^{m, h+1}_{u} = \textsc{Drop}(\widetilde{\mathbf{H}}_u^m) \cdot \textsc{Drop}(\widetilde{\mathbf{H}}_i^{m \top}) \cdot \mathbf{E}^{m,h}_{i}.
\end{equation}
Apparently, the hypergraph passing explicitly enables global information transfer by taking the item collaborative embedding and modality-aware hypergraph dependencies as input. Then, we can obtain the global embeddings matrix $\mathbf{E}_{ghe}$ by aggregating global embeddings from all modalities,
\begin{equation}
\mathbf{E}_{ghe} = \sum\limits_{m \in \mathcal{M}} \mathbf{E}^{m, H}, \quad \mathbf{E}^{m, H}=[\mathbf{E}^{m, H}_u,\mathbf{E}^{m, H}_i],
\end{equation}
where $\mathbf{E}^{m, H}_{u} \in \mathbb{R}^{|\mathcal{U}|\times d}$ and $\mathbf{E}^{m, H}_{i} \in \mathbb{R}^{|\mathcal{I}|\times d}$ are global embedding matrices of user $u$ and item $i$ obtained in the $H$-th hypergraph layer under modality $m$, respectively.

To further achieve the robust fusion of global embeddings among different modalities, we develop cross-modal hypergraph contrastive learning to distill the self-supervision signals for global interest consistency. Specifically, we take the global embeddings of users acquired in different modalities as positive pairs and different users as negative pairs, and then employ the InfoNCE~\cite{gutmann2010noise} to formally define user-side hypergraph contrastive loss as,
\begin{equation}
	\mathcal{L}^u_{\mathrm{HCL}} = \sum_{u \in \mathcal{U}} - \log 
	\frac{\exp(s(\mathbf{E}^{v, H}_{u}, \mathbf{E}^{t, H}_{u}) / \tau)}
	{\sum_{u' \in \mathcal{U}} \exp(s(\mathbf{E}^{v, H}_{u}, \mathbf{E}^{t, H}_{u'})/ \tau)},
\end{equation}
where $s(\cdot)$ is the cosine function, and $\tau$ is the temperature factor, generally set to $0.2$. Note here we only consider visual and textual modalities, i.e., $m\in \{v,t\}$. Similarly, we can define item-side cross-modal contrastive loss $\mathcal{L}^i_{\mathrm{HCL}}$.

\subsection{Fusion and Prediction}
We acquire the final representations $\mathbf{E}^*$ of users and items by fusing their two types of local embeddings $\mathbf{E}_{lge}^{id}$, $\mathbf{E}_{lge}^{m}$ and global embeddings $\mathbf{E}_{ghe}$,
\begin{equation}
	\mathbf{E}^* =\mathbf{E}_{lge}^{id} + \sum_{m \in \mathcal{M}} \textsc{Norm}(\mathbf{E}_{lge}^{m}) + \alpha \cdot \textsc{Norm}(\mathbf{E}_{ghe}),
\end{equation}
where $\textsc{Norm}(\cdot)$ is a normalization function to alleviate the value scale difference among embeddings, $\alpha$ is an adjustable factor to control the integration of global embeddings. 

We then use inner product to calculate the preference score $\hat{r}_{u,i}$ of user $u$ towards item $i$, i.e., $\hat{r}_{u,i}=\mathbf{e}^*_u \cdot {\mathbf{e}^*_i}^{\top}$. The Bayesian personalized ranking (BPR) loss~\cite{rendle2012bpr} is employed to optimize model parameters,
\begin{equation}
	\mathcal{L}_{\mathrm{BPR}}=-\sum_{(u,i^{+},i^{-}) \in \mathcal{R}} \ln \sigma\left(\hat{r}_{u,i^{+}}-\hat{r}_{u, i^{-}}\right) + \lambda_1 \| \mathbf{\Theta}\|_2^2,
\end{equation}
where $\mathcal{R}=\{(u, i^{+}, i^{-}) | (u, i^{+}) \in \mathcal{G}, (u, i^{-}) \notin \mathcal{G}\}$ is a set of triples for training,  $\sigma(\cdot)$ is the sigmoid function, and $\lambda_1$ and $\mathbf{\Theta}$ represent the regularization coefficient and model parameters, respectively.

Finally, we integrate hypergraph contrastive loss with the BPR~\cite{rendle2012bpr} loss into a unified objective as,
\begin{equation}
	\mathcal{L} = \mathcal{L}_{\mathrm{BPR}} + \lambda_2 \cdot (\mathcal{L}^u_{\mathrm{HCL}} + \mathcal{L}^i_{\mathrm{HCL}})
\end{equation}
where $\lambda_2$ is a hyperparameter for loss term weighting. We minimize the joint objective $\mathcal{L}$ by using Adam optimizer~\cite{kingma2014adam}. The weight-decay regularization term is applied over model parameters $\mathbf{\Theta}$.

	\section{Experiment}
	\label{sec:experiment}

\setlength{\tabcolsep}{1.5mm}
\begin{table}[t]
	\centering
	\begin{tabular}{lrrrr}
		\hline
		\hline
		\textbf{Dataset} & \textbf{\#User} & \textbf{\#Item} & \textbf{\#Interaction} & \textbf{Sparsity}\\
		\hline
		\hline
		\textbf{Baby} & 19,445 & 7,050 & 160,792  & 99.883\% \\
		\textbf{Sports} & 35,598 & 18,357 & 296,337  & 99.955\% \\
		\textbf{Clothing} & 39,387 & 23,033 & 278,677  & 99.969\% \\
		\hline
		\hline
	\end{tabular}%
    \caption{Statistics of the three evaluation datasets}
	\label{tab:datasets}%
\end{table}

\setlength{\tabcolsep}{0.85mm}
\begin{table*}[t]
	\centering
	\begin{tabular}{l|cccc|cccc|cccc}
		\hline
		\hline
		\textbf{Datasets} & \multicolumn{4}{c|}{\textbf{Baby}} & \multicolumn{4}{c|}{\textbf{Sports}} & \multicolumn{4}{c}{\textbf{Clothing}} \\
		\hline
		\textbf{Metrics} & \textbf{R@10} & \textbf{R@20} & \textbf{N@10} & \textbf{N@20} & \textbf{R@10} & \textbf{R@20} & \textbf{N@10} & \textbf{N@20} & \textbf{R@10} & \textbf{R@20} & \textbf{N@10} & \textbf{N@20} \\
		\hline
		\hline
		\textbf{BPR} & 0.0379  & 0.0607  & 0.0202  & 0.0261  & 0.0452  & 0.0690  & 0.0252  & 0.0314  & 0.0211  & 0.0315  & 0.0118  & 0.0144  \\
		\hline
		\textbf{LightGCN} & 0.0464  & 0.0732  & 0.0251  & 0.0320  & 0.0553  & 0.0829  & 0.0307  & 0.0379  & 0.0331  & 0.0514  & 0.0181  & 0.0227  \\
		\textbf{SGL} & 0.0532  & 0.0820  & 0.0289  & 0.0363  & 0.0620  & 0.0945  & 0.0339  & 0.0423  & 0.0392  & 0.0586  & 0.0216  & 0.0266  \\
		\textbf{NCL} & 0.0538  & 0.0836  & 0.0292  & 0.0369  & 0.0616  & 0.0940  & 0.0339  & 0.0421  & 0.0410  & 0.0607  & 0.0228  & 0.0275  \\
		\hline
		\textbf{HCCF} & 0.0480  & 0.0756  & 0.0259  & 0.0332  & 0.0573  & 0.0857  & 0.0317  & 0.0394  & 0.0342  & 0.0533  & 0.0187  & 0.0235  \\
		\textbf{SHT} & 0.0470 & 0.0748  & 0.0256  & 0.0329  & 0.0564  & 0.0838  & 0.0306  & 0.0384  & 0.0345 & 0.0541 & 0.0192 & 0.0243  \\
		\hline
		\textbf{VBPR} & 0.0424  & 0.0663  & 0.0223  & 0.0284  & 0.0556  & 0.0854  & 0.0301  & 0.0378  & 0.0281  & 0.0412  & 0.0158  & 0.0191  \\
		\textbf{MMGCN} & 0.0398  & 0.0649  & 0.0211  & 0.0275  & 0.0382  & 0.0625  & 0.0200  & 0.0263  & 0.0229  & 0.0363  & 0.0118  & 0.0152  \\
		\textbf{GRCN} & 0.0531  & 0.0835  & 0.0291  & 0.0370  & 0.0600  & 0.0921  & 0.0324  & 0.0407  & 0.0431  & 0.0664  & 0.0230  & 0.0289  \\
		\textbf{LATTICE} & 0.0536  & 0.0858  & 0.0287  & 0.0370  & 0.0618  & 0.0950  & 0.0337  & 0.0423  & 0.0459 & 0.0702 & 0.0253 & 0.0306 \\
		\textbf{MMGCL} & 0.0522  & 0.0778  & 0.0289  & 0.0355  & 0.0660  & 0.0994  & 0.0362  & 0.0448  & 0.0438  & 0.0669  & 0.0239  & 0.0297  \\
		\textbf{MICRO} & \underline{0.0570}  & \underline{0.0905}  &\underline{0.0310}  & \underline{0.0406}  & 0.0675  & \underline{0.1026}  & 0.0365  & \underline{0.0463}  & \underline{0.0496} & \underline{0.0743}  & \underline{0.0264}  & \underline{0.0332}  \\
		\textbf{SLMRec} & 0.0540  & 0.0810  & 0.0296  & 0.0361  & \underline{0.0676}  & 0.1007  & \underline{0.0374}  & 0.0462  & 0.0452  & 0.0675  & 0.0247  & 0.0303  \\
		\textbf{BM3} & 0.0538  & 0.0857  & 0.0301  & 0.0378  & 0.0659  & 0.0979  & 0.0354  & 0.0437  & 0.0450  & 0.0669  & 0.0243  & 0.0295  \\
		\hline
		\textbf{LGMRec} & \textbf{0.0644*} & \textbf{0.1002*} & \textbf{0.0349*} & \textbf{0.0440*} & \textbf{0.0720*} & \textbf{0.1068*} & \textbf{0.0390*} & \textbf{0.0480*} & \textbf{0.0555*} & \textbf{0.0828*} & \textbf{0.0302*} & \textbf{0.0371*} \\
		\textbf{Improv.} & 12.98\% & 10.72\% & 12.58\% & 8.37\% & 6.51\% & 4.09\% & 4.28\% & 3.67\% & 11.90\% & 11.44\% & 14.39\% & 1.75\% \\
		
		\hline
		\hline
	\end{tabular}%
    \caption{Overall performances of LGMRec and other baselines on three datasets. The best result is in boldface and the second best is underlined. The t-tests validate the significance of performance improvements with $p$-value $\le 0.05$.}
	\label{tab:comparison}%
\end{table*}%

\subsection{Experimental Settings}
\subsubsection{Datasets}
To evaluate our proposed model, we conduct comprehensive experiments on three widely used Amazon datasets~\cite{mcauley2015image}: Baby, Sports and Outdoors, Clothing Shoes and Jewelry. We refer to them as \textbf{Baby}, \textbf{Sports}, \textbf{Clothing} for brief. We adopt the $5$-core setting to filter users and items for each dataset. The three datasets include both visual and textual modal features. In this work, we use the $4096$-dimensional original visual features and $384$-dimensional original textual features that have been extracted and published in prior work~\cite{zhou2023bootstrap}. The statistics of the three datasets are summarized in Table~\ref{tab:datasets}. 

\subsubsection{Evaluation Protocols}
For each dataset, we randomly split historical interactions into training, validation, and testing sets with $8:1:1$ ratio. 
Two widely used protocols are used to evaluate the performance of top-$n$ recommendation: Recall (R@$n$) and Normalized Discounted Cumulative Gain~\cite{he2015trirank} (N@$n$). We tune $n$ in $\{10, 20\}$ and report the average results for all users in the testing set.

\subsubsection{Parameter Settings}
For a fair comparison, we optimize all models with the default batch size $2048$, learning rate $0.001$, and embedding size $d=64$. For all graph-based methods, the number $L$ of collaborative graph prorogation layers is set to $2$. In addition, we initialize the model parameters with the Xavier method~\cite{glorot2010understanding}. For our model, the optimal hyper-parameters are determined via grid search on the validation set. Specifically, the number of modal graph embedding layers and hypergraph embedding layers ($K$ and $H$) are tuned in $\{1,2,3,4\}$. The number $A$ of hyperedge is searched in $\{1,2,4,8,16,32,64,128,256\}$. The dropout ratio $\rho$ and the adjust factor $\alpha$ are tuned in $\{0.1, 0.2, \ldots, 1.0\}$. We search both the adjust weight $\lambda_2$ of contrastive loss and the regularization coefficient $\lambda_1$ in $\{1e^{-6}, 1e^{-5}, \ldots, 0.1\}$. The early stop mechanism is adopted, i.e., the training will stop when R@$20$ on the verification set does not increase for $20$ successive epochs. We implement LGMRec\footnote{https://github.com/georgeguo-cn/LGMRec} with MMRec~\cite{zhou2023mmrec}.

\subsubsection{Baselines}
We compare our proposed LGMRec with the following four groups of recommendation baselines, including (1) General CF Models: \textbf{BPR}~\cite{rendle2012bpr}; (2) Graph-based Recommenders: \textbf{LightGCN}~\cite{he2020lightgcn}, \textbf{SGL}~\cite{wu2021self}, \textbf{NCL}~\cite{lin2022improving}; (3) Hypergraph-based Recommenders: \textbf{HCCF}~\cite{xia2022hypergraph}, \textbf{SHT}~\cite{xia2022self}; and (4) Multi-Modal Recommenders: \textbf{VBPR}~\cite{he2016vbpr}, \textbf{MMGCN}~\cite{wei2019mmgcn}, \textbf{GRCN}~\cite{wei2020graph}, \textbf{LATTICE}~\cite{zhang2021mining}, \textbf{MMGCL}~\cite{yi2022multi}, \textbf{MICRO}~\cite{zhang2022latent} \textbf{SLMRec}~\cite{tao2022self}, \textbf{BM3}~\cite{zhou2023bootstrap}.

\subsection{Performance Comparison}
The performance comparison for all methods on the three datasets is summarized in Table~\ref{tab:comparison}, from which we have the following key observations: (1) \textbf{The superiority of LGMRec}. LGMRec substantially outperforms all other baselines and achieves promising performance across different datasets. We attribute such significant improvements to: i) The modeling of separated local embeddings that excavates user decoupled interests; ii) The hypergraph learning injects the modality-related global dependencies to local graph embeddings to mitigate interactive sparsity. (2) \textbf{The effectiveness of modal features}. Introducing knowledge-rich modality information is beneficial for boosting performance. Experimentally, though only linearly fusing the ID embeddings and modal features of items, the performance of VBPR still outperforms its counterpart (i.e., BPR). By effectively modeling the modal information, the multimodal recommenders (e.g., LATTICE, SLMRec, BM3) with LightGCN as the backbone network basically achieve better results than LightGCN. (3) \textbf{The effectiveness of hypergraph learning}. Hypergraph-based recommenders (i.e., HCCF and SHT) outperform the graph-based CF model LightGCN, suggesting the effectiveness of modeling global dependencies under hypergraph architecture. Besides, the significant improvement of LGMRec over  competitive baselines further demonstrates the potential of hypergraph networks in modeling modality-aware global dependencies.

\setlength{\tabcolsep}{0.5mm}
\begin{table}[t]
	\centering
	\begin{tabular}{l|cc|cc|cc}
		\hline
		\hline
		\textbf{Components} & \multicolumn{2}{c|}{\textbf{Baby}} & \multicolumn{2}{c|}{\textbf{Sports}} & \multicolumn{2}{c}{\textbf{Clothing}} \\
		\hline
		\textbf{Metrics} & \textbf{R@20} & \textbf{N@20} & \textbf{R@20} & \textbf{N@20} & \textbf{R@20} & \textbf{N@20} \\
		\hline
		\hline
		\textit{w/o MM}  & 0.0732  & 0.0320  & 0.0829  & 0.0379  & 0.0514  & 0.0227  \\
		\hline
		\textit{w/o LGE} & 0.0806  & 0.0351  & 0.0851  & 0.0392  & 0.0741  & 0.0327  \\
		\textit{w/o CGE} & 0.0947  & 0.0423  & 0.0997  & 0.0448  & 0.0807  & 0.0360  \\
		\textit{w/o MGE}  & 0.0929  & 0.0417  & 0.0988  & 0.0440  & 0.0804  & 0.0357  \\
		\hline
		\textit{w/o GHE} & 0.0972  & 0.0430  & 0.1032  & 0.0468  & 0.0803  & 0.0364  \\
		\textit{w/o HCL} & 0.0992  & 0.0434  & 0.1051  & 0.0474  & 0.0812  & 0.0368  \\
		\hline
		\textit{w/ SUID}  & 0.0869  & 0.0379  & 0.0895  & 0.0395  & 0.0713  & 0.0307  \\
		\hline
		\textbf{LGMRec} & \textbf{0.1002} & \textbf{0.0440} & \textbf{0.1068} & \textbf{0.0480} & \textbf{0.0828} & \textbf{0.0371} \\
		\hline
		\hline
	\end{tabular}%
	\caption{Ablation of different components on LGMRec.}
	\label{tab:ablation}%
\end{table}%

\subsection{Ablation Study}
We conduct ablation studies to explore the compositional effects of LGMRec. From the results reported in Table~\ref{tab:ablation}, we can find: (1) The variant \textit{w/o MM} without multimodal contents degenerates into LightGCN and achieves the worst performance, indicating that introducing modality features can greatly improve accuracy. (2) Removing either \textit{LGE} or \textit{GHE} can cause performance drops of LGMRec, demonstrating the benefits of modeling both local and global user interests. Notably, the variant \textit{w/o LGE} performs worse than \textit{w/o GHE}, which indicates that local interests directly related to user behavior are more important, and global interests can serve as a supplement. (3) In local graph embeddings, the variant \textit{w/o CGE} (with \textit{MGE} only) achieves better performance than \textit{w/o MGE} (with \textit{CGE} only) on all datasets, which reveals the importance of integrating multimodal features into user-item interaction modeling. (4) The variant \textit{w/o HCL} removes hypergraph contrastive learning and only linearly adds all global embeddings. Its performances indicate that contrastive fusion of global embeddings of different modalities can improve performance by modeling the inter-modal global semantic consistency. (5) The variant \textit{w/ SUID} that still shares user ID embeddings in both MGE and CGE modules  performs worse than LGMRec, verifying the benefits of independently modeling user decoupled interests.

\begin{figure}[t]
	\centering
	\includegraphics[width=0.49\linewidth]{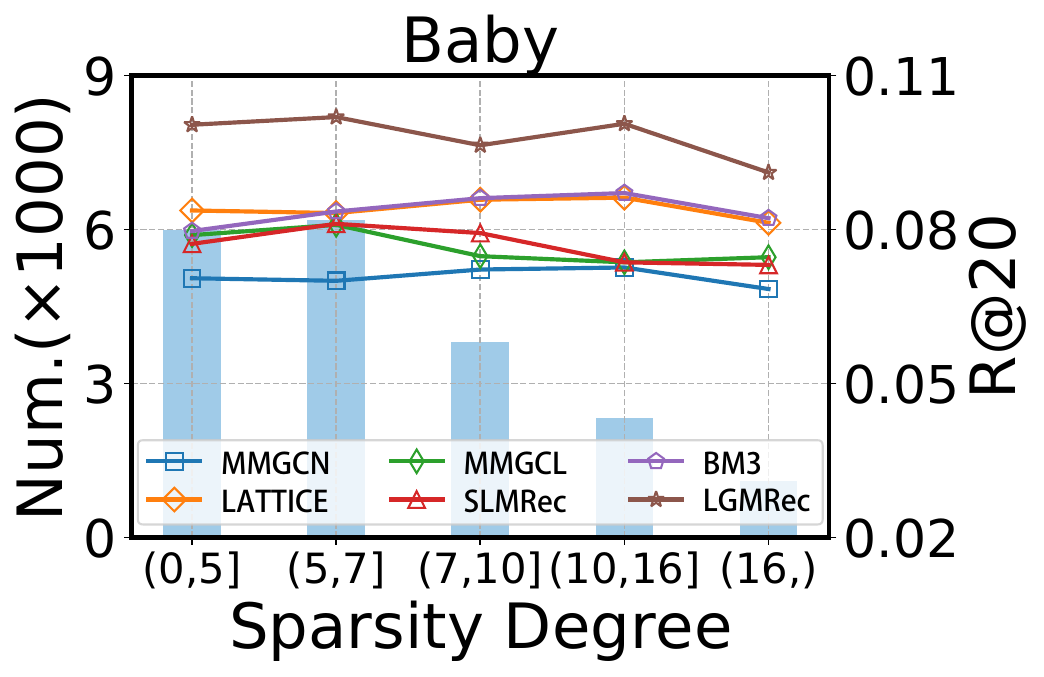}
	\includegraphics[width=0.49\linewidth]{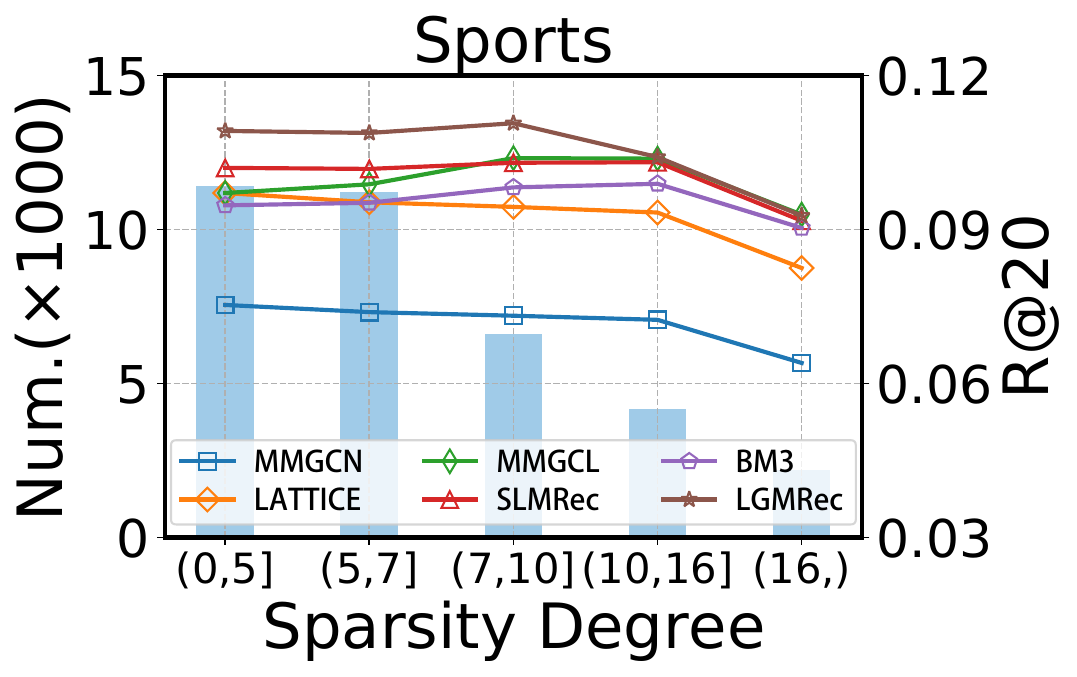}\\
	\caption{Performance w.r.t. different user interaction sparsity degrees	in terms of R@$20$ on Baby and Sports datasets.}
	\label{fig:sparse}
\end{figure}

\begin{figure}[t]
	\centering
	\includegraphics[width=0.49\linewidth, height=0.32\linewidth]{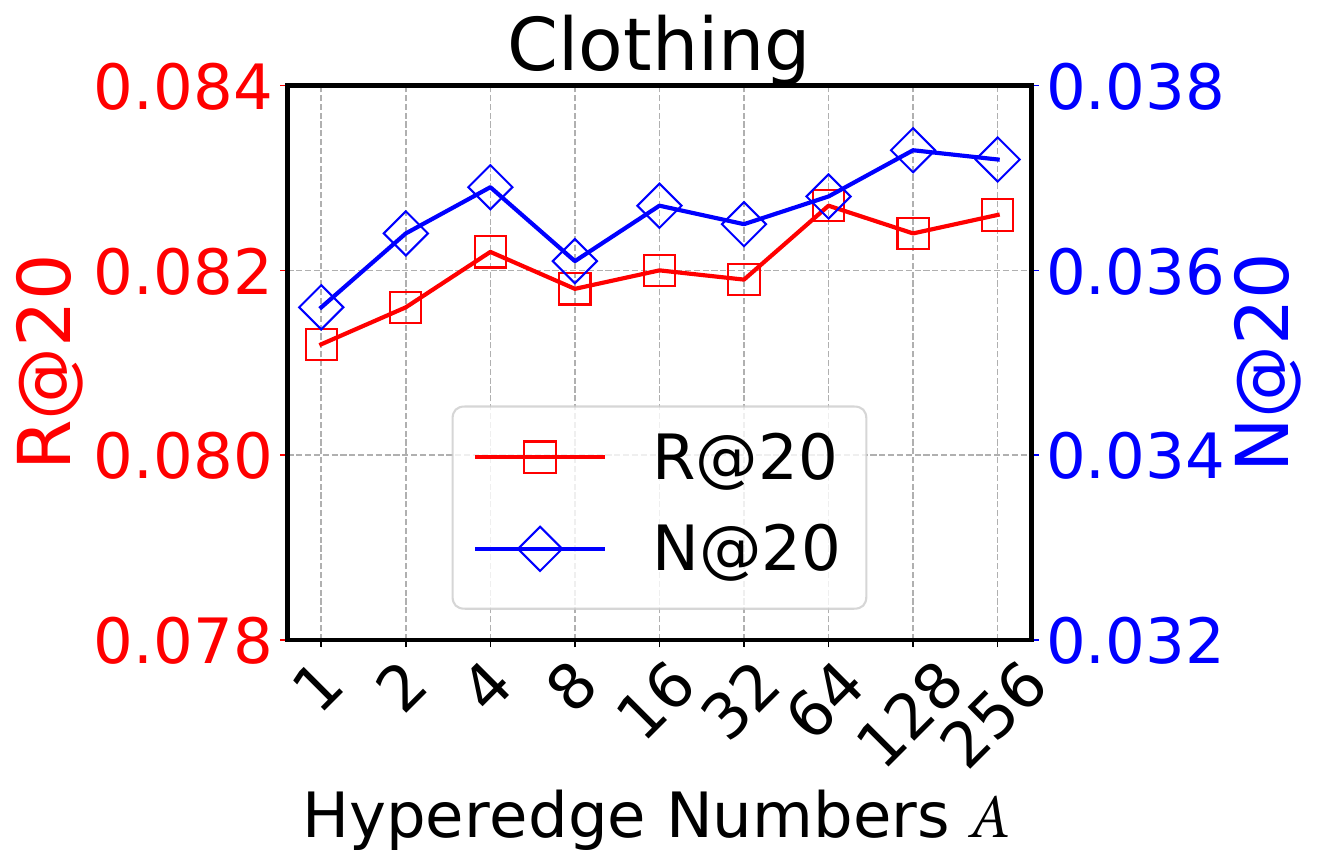}
	\includegraphics[width=0.49\linewidth, height=0.32\linewidth]{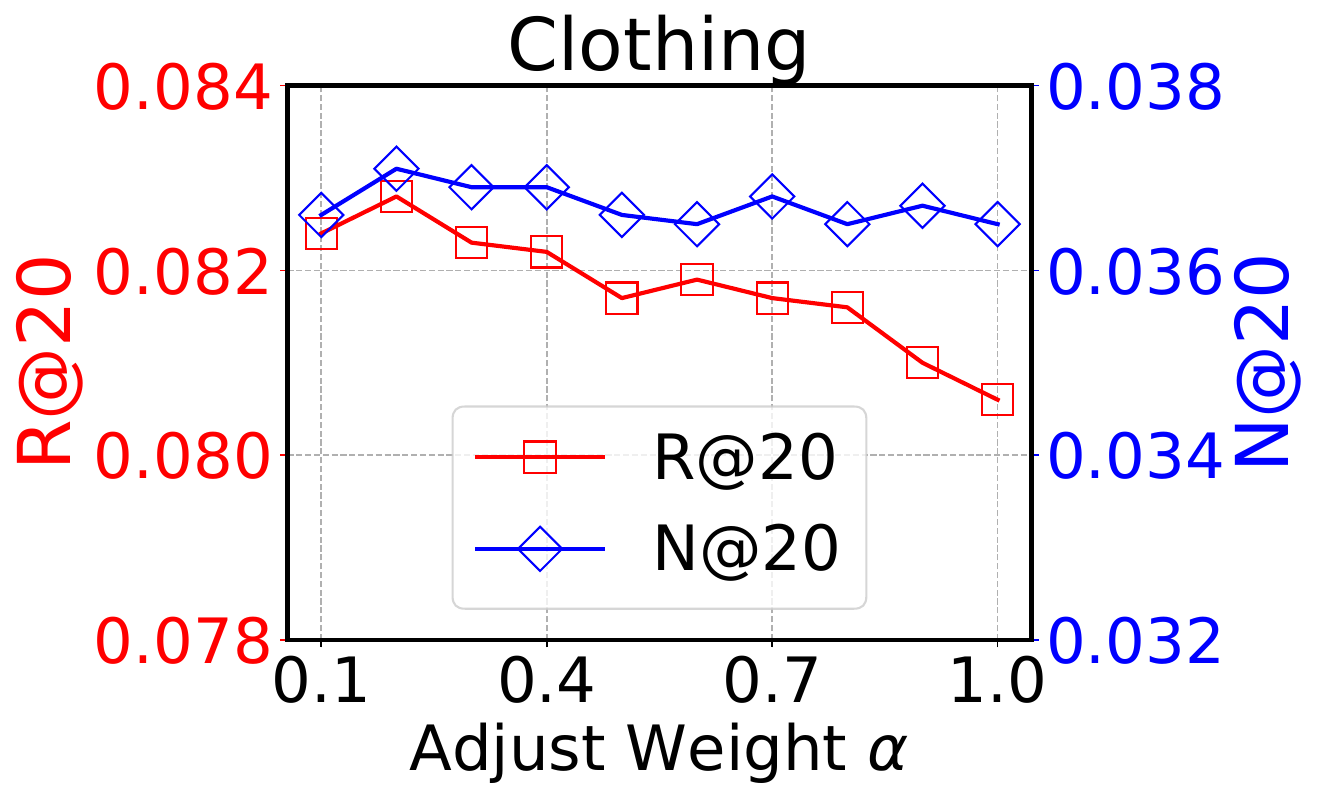}\\
	\caption{Performances under different settings of two key hyperparameters ($A$ and $\alpha$) on Clothing datasets.}
	\label{fig:parameters}
\end{figure}

\subsection{In-Depth Analysis}
\subsubsection{\textbf{Performance with Different Data Sparsity}}
We further study the influence of sparse user interactions by comparing LGMRec with five representative multimodal recommendation baselines: MMGCN, LATTICE, MMGCL, SLMRec, and BM3, on Baby and Sports datasets. Multiple user groups are constructed according to the number of their interactions. For example, the first user group contains users interacting with $0-5$ items. From the results in Figure~\ref{fig:sparse}, we can observe that: (1) The superior performance of LGMRec is consistent across user groups with different sparsity degrees, revealing the effectiveness of LGMRec in alleviating interaction sparsity by modeling local and global representations. (2) LGMRec achieves more performance gains on sparser user groups. Specifically, LGMRec realizes 19.95\% and 10.83\% improvements over the best baseline for the sparsest and densest group on Baby, respectively, verifying the robustness of LGMRec in dealing with sparser user interactions.

\begin{figure}[t]
	\centering
	\includegraphics[width=0.95\linewidth]{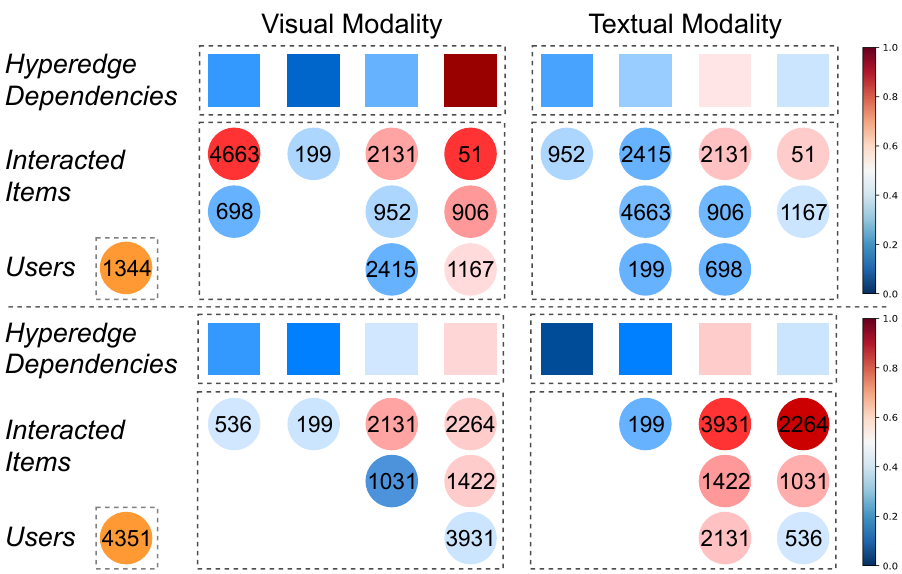}\\
	\caption{Case study of learned global dependencies of two users $u_{1344}$ and $u_{4351}$ with four hyperedges on Baby dataset.}
	\label{fig:case}
\end{figure}

\subsubsection{Hyperparameter Analysis}
\label{hyperparameter}
Figure~\ref{fig:parameters} reports the impact of two key hyperparameters of LGMRec on Clothing dataset:
\paragraph{Hyperedge number $A$.} From the left figure in Figure~\ref{fig:parameters}, we can observe that LGMRec presents performance promotion as the number of hyperedges increases, demonstrating the effectiveness of capturing multi-hyperedge global structures, especially for sparser Clothing datasets. 

\paragraph{Adjustable weight $\alpha$.} Impact of weight $\alpha$ of fusing global embeddings is also investigated in Figure~\ref{fig:parameters}. We can find that the performance first rises to an optimal value ($\alpha = 0.2$) and then declines, which suggests that an appropriate $\alpha$ can improve accuracy by properly supplementing global embeddings, but a too large $\alpha$ may negatively affect performance.

\subsection{Case Study}
We qualitatively study the global hypergraph dependencies. Specifically, we randomly select two users $u_{1344}$, $u_{4351}$ with similar global embeddings learned on Baby dataset. Hypergraph dependencies under visual and textual modalities for the two users and the items they interact with are presented in Figure~\ref{fig:case}. The four hyperedges (squares) are shaded depending on the user-hyperedge dependency score. Moreover, the interacted items (circles) are arranged below the corresponding hyperedges in order, according to the maximum item-hyperedge dependency score. From Figure~\ref{fig:case}, we can observe that: 
(1) The user-hyperedge dependencies differ in different modalities. For example, the global interests of user $u_{1344}$ in the visual modality are mainly related to the $4$-th attribute hyperedge. Under the textual modality, user $u_{1344}$ has larger dependency scores with the $3$-rd hyperedges. Thus, we guess that the four items ($i_{51}$, $i_{906}$, $i_{1167}$, and $i_{2131}$) closely related to head hyperedges can reflect user $u_{1344}$'s true preferences, while item $i_{4663}$ attached to the $1$-st hyperedge may be a noise interaction. (2) Although the interacted items are largely non-overlapping, user $u_{4351}$ and user $u_{1344}$ still have similar hyperedge dependencies, demonstrating why their global embeddings are similar. The results further reveal that LGMRec can exploit global hypergraph learning to distill similar knowledge of item modal features for performance improvement.

	\section{Conclusion}
	\label{sec:conclusion}

In this work, we proposed a novel model LGMRec for MRSs, which captures and utilizes local embeddings with local topological information and global embeddings with hypergraph dependencies. Specifically, we adopted a local graph embedding module to independently learn collaborative-related and modality-related local user interests. A global hypergraph embedding module is further designed to mine global user interests. Extensive experiments on three datasets demonstrated the superiority of our model over various baselines. For future work, we intend to seek better means of modeling the differences and commonalities among modalities for further performance improvement.
	
	\section{Acknowledgements}
	We would like to thank all anonymous reviewers for their valuable comments. The work was partially supported by the National Key R\&D Program of China under Grant No. 2022YFC3802101 and the National Natural Science Foundation of China under Grant No. 62272176.
	
	\bibliography{aaai24}
	
	\clearpage

\appendix

\section{Appendix}

\subsection{Complexity Analysis of LGMRec}
We conduct the complexity analysis of LGMRec. The computational cost of LGMRec mainly comes from two parts. In the local graph embedding module, the collaborative graph embedding has a $\mathcal{O}(L \times |\mathcal{E}| \times d)$ complexity, where $L$ is the number of graph message passing layers, $|\mathcal{E}|$ is the number of edges in user-item interaction graph $\mathcal{G}$. For modality graph embedding, the computational cost of modal feature initialization is $\mathcal{O}(|\mathcal{M}| \times d_m \times d)$, where $|\mathcal{M}|$ is the number of modalities. The modality graph propagation has the same computational complexity as the collaborative graph embedding. Thus, the overall time complexity of the local graph embedding module is $\mathcal{O}(((L+K) \times |\mathcal{E}| + |\mathcal{M}| \times d_m) \times d)$.  

For the global hypergraph embedding module, the time complexity of hypergraph dependency constructing is $\mathcal{O}(|\mathcal{M}| \times A \times |\mathcal{I}| \times (|\mathcal{U}|+ d_m))$, where $A$ is the number of hyperedges. The hypergraph message passing schema takes $\mathcal{O}(|\mathcal{M}| \times (|\mathcal{I}| \times H + |\mathcal{U}|) \times A \times d)$ complexity with the global information propagation, where $H$ is the number of hypergraph layers. The cost of the hypergraph contrastive learning is $\mathcal{O}(B \times (|\mathcal{U}| + |\mathcal{I}|) \times d)$ with only two modalities $v$ and $t$, where $B$ is the batch size. The overall time complexity of the global hypergraph embedding module is $\mathcal{O}(|\mathcal{M}| \times A \times (|\mathcal{I}| \times ((|\mathcal{U}|+ d_m) + H \times d) + |\mathcal{U}|))$. In practice, our two modules can be executed in parallel, which makes LGMRec quite efficient in actual execution. During the training process, its actual running time is comparable to existing methods, such as MMGCL~\cite{yi2022multi}, and SLMRec~\cite{tao2022self}. Compared to existing graph-based multimodal recommenders, LGMRec only involves $\mathcal{O}(|\mathcal{M}| \times A \times d_m)$ extra parameters for the memory cost.

\subsection{Baselines}
\noindent
\textbf{(i) General CF Models}
\begin{itemize}[leftmargin=1em]
	\item \textbf{BPR}~\cite{rendle2012bpr} maps the user and item in a low-dimensional latent embedding space and utilizes Bayesian pairwise ranking loss to optimize model parameters.
\end{itemize}

\noindent
\textbf{(ii) Graph-based Recommendations}
\begin{itemize}[leftmargin=1em]
	\item \textbf{LightGCN}~\cite{he2020lightgcn} is a typical graph-based CF method that utilizes light graph convolutional networks to learn high-order connectivity of users and items.
	
	\item \textbf{SGL}~\cite{wu2021self} introduces contrastive learning to enhance graph collaborative filtering. We implement this method by data augmentation with random edge dropout.
	
	\item \textbf{NCL}~\cite{lin2022improving} enhances the graph-based CF model by identifying structural and semantic neighboring nodes as positive samples to construct contrastive views.
\end{itemize}

\noindent
\textbf{(ii) Hypergraph-based Recommendations}
\begin{itemize}[leftmargin=1em]
	\item \textbf{HCCF}~\cite{xia2022hypergraph} leverages the hypergraph neural network to inject the global collaborative relations into the graph-based recommendation. 
	
	\item \textbf{SHT}~\cite{xia2022self} captures the global collaborative embeddings for contrastive learning by joint utilizing hypergraph encoder and multi-head attention mechanism.
\end{itemize}

\noindent
\textbf{(iii) Multi-Modal Recommendations}
\begin{itemize}[leftmargin=1em]
	\item \textbf{VBPR}~\cite{he2016vbpr} integrates modal features with ID embeddings to extend the traditional CF paradigm.
	
	\item \textbf{MMGCN}~\cite{wei2019mmgcn} learns fine-grained modality-specific user preferences by achieving message-passing on the user-item bipartite graph of each modality.
	
	\item \textbf{GRCN}~\cite{wei2020graph} is a structure-refined graph multimedia recommender, in which modality contents are used to adjust the structure of interaction graph by identifying the noisy edges.
	
	\item \textbf{LATTICE}~\cite{zhang2021mining} exploits multi-modal features to mine the latent semantic structure between items to improve multi-modal recommendation.
	
	\item \textbf{MMGCL}~\cite{yi2022multi} includes contrastive learning into multimodal recommendation via graph augmentation with modality-related edge dropout and masking.
	
	\item \textbf{MICRO}~\cite{zhang2022latent} extends the LATTICE~\cite{zhang2021mining} to fuse multimodal features by introducing contrastive learning to capture modality-shared and modality-specific information.
	
	\item \textbf{SLMRec}~\cite{tao2022self} devises three types of data augmentation at different granularity to achieve multi-modal self-supervised tasks.
	
	\item \textbf{BM3}~\cite{zhou2023bootstrap} utilizes a simple latent embedding dropout mechanism to generate contrastive view in self-supervised learning for multimodal recommendation.
\end{itemize}

\begin{table}[t]
	\centering
	\setlength{\tabcolsep}{1.5mm}{
		\begin{tabular}{l|cccccccc}
			\hline
			\hline
			\textbf{Parameters} & \textbf{$L$} & \textbf{$K$} & \textbf{$H$} & \textbf{$A$} & \textbf{$\alpha$} & \textbf{$\rho$} & \textbf{$\lambda_1$} & \textbf{$\lambda_2$} \\
			\hline
			\textbf{Baby} & 2 & 2 &  1 & 4 &  0.3& 0.5& $1e^{-6}$ & $1e^{-4}$ \\
			\textbf{Sports} & 4 & 2 &  1 & 4 &  0.6& 0.4& $1e^{-6}$ & $1e^{-4}$\\
			\textbf{Clothing} & 3 & 2 &  2 & 64 &  0.2& 0.2& $1e^{-6}$ & $1e^{-4}$\\
			\hline
			\hline
	\end{tabular}}
	\caption{Parameter setting for three datasets.}
	\label{tab:paramter_set}
\end{table}

\subsection{Parameter Setting}
For a fair comparison, we optimize all models with the default batch size $2048$, learning rate $0.001$, and embedding size $d=64$. Table~\ref{tab:paramter_set} presents other optimal parameter settings for the three datasets. In addition, all experiments in this paper are performed in the same experimental environment with Intel(R) Xeon(R) Silver 4210R CPU @ 2.40GHz and GeForce RTX 3090.

\begin{figure*}[t]
	\centering
	\subfloat[\scriptsize{Impact of collaborative graph layers $L$.}]{
		\includegraphics[width=0.16\linewidth]{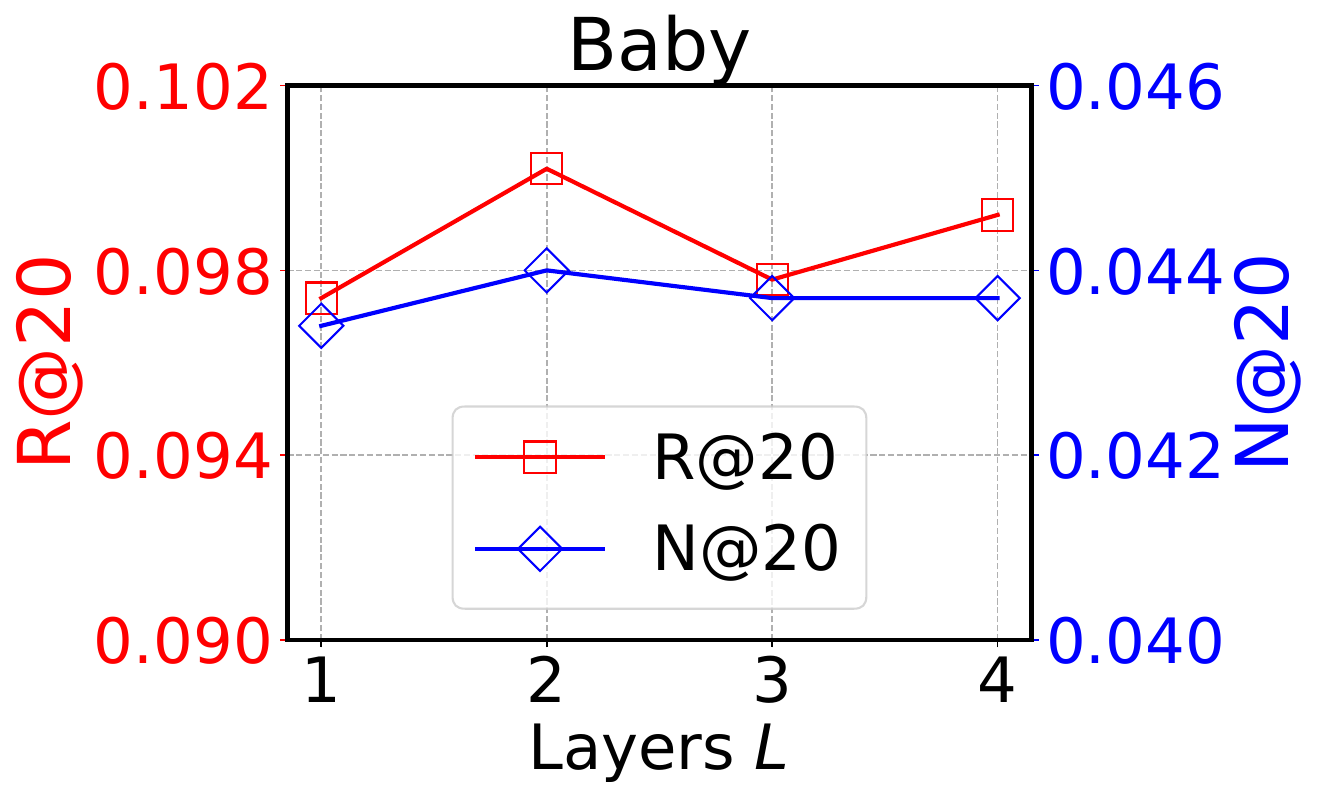}
		\includegraphics[width=0.16\linewidth]{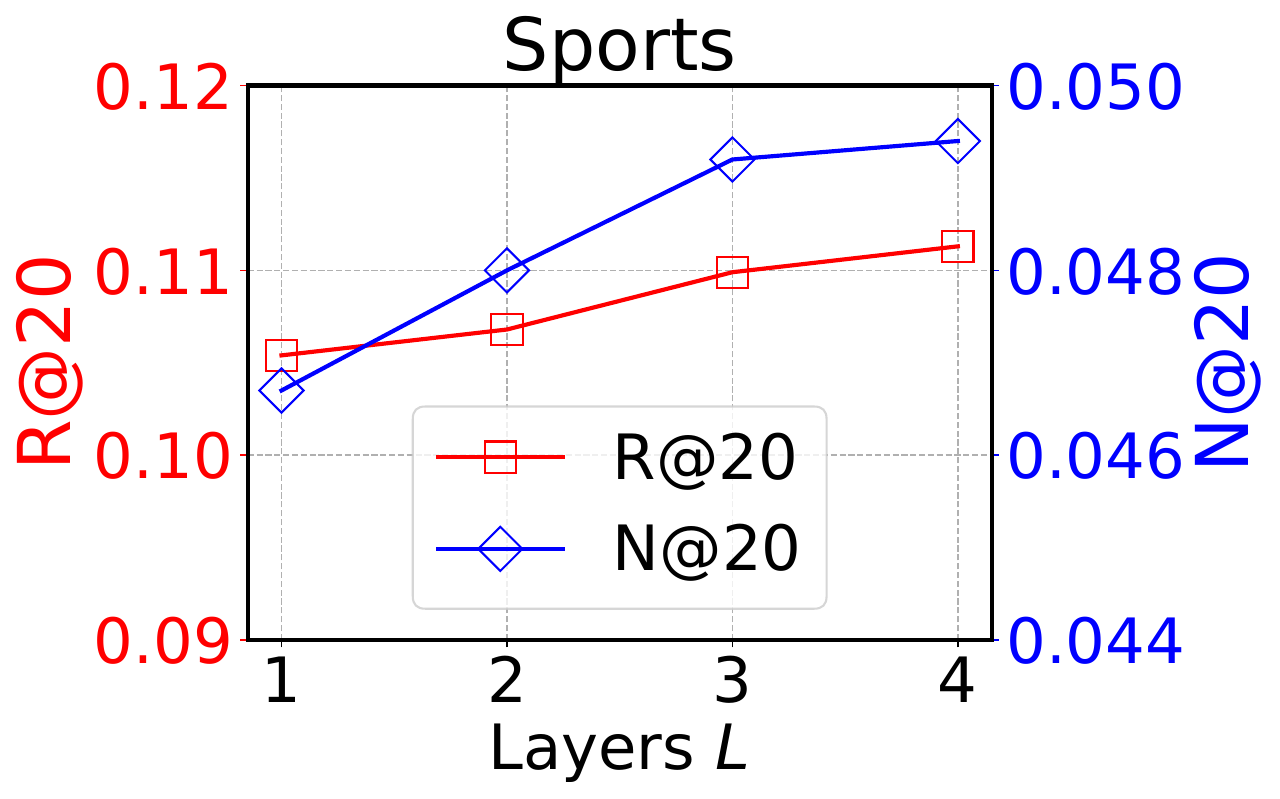}
		\includegraphics[width=0.16\linewidth]{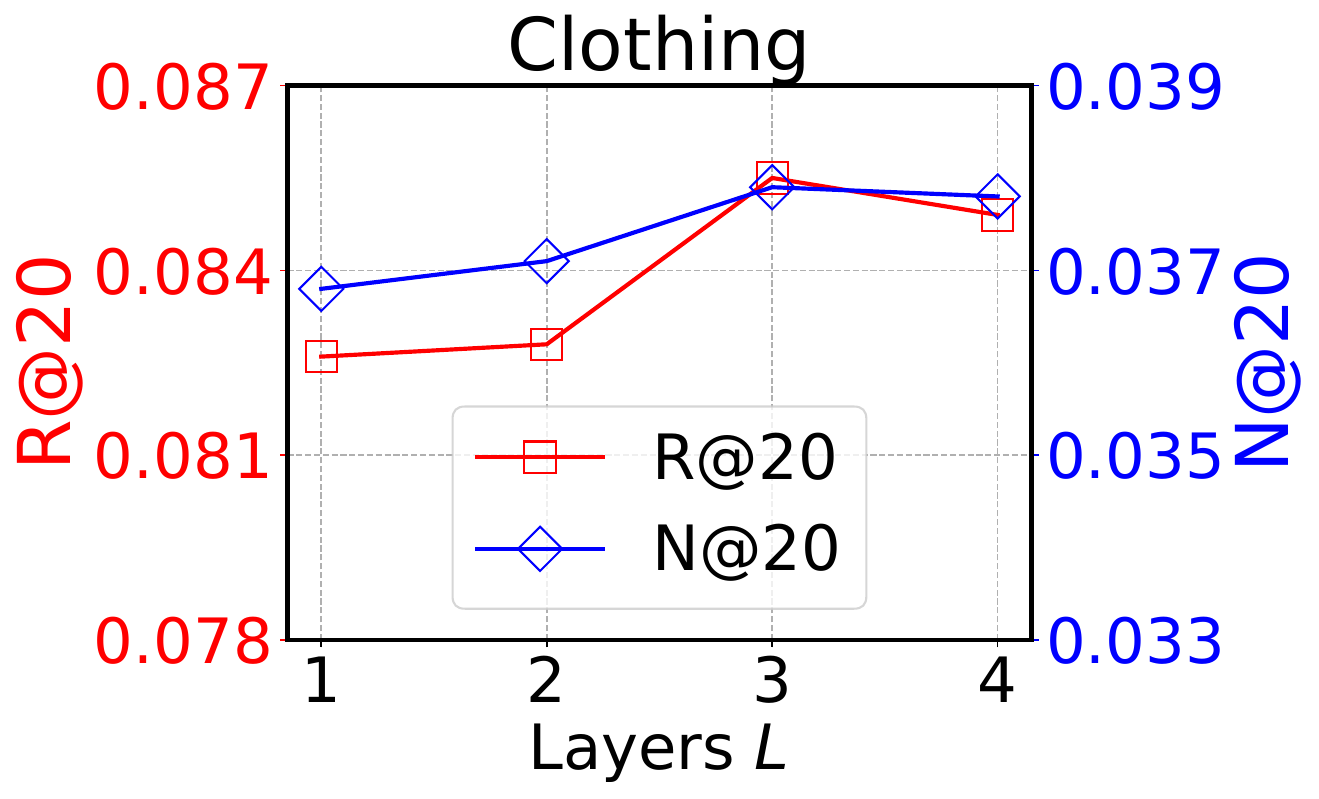}
	}
    \subfloat[\scriptsize{Impact of modality graph layers $K$.}]{
		\includegraphics[width=0.16\linewidth]{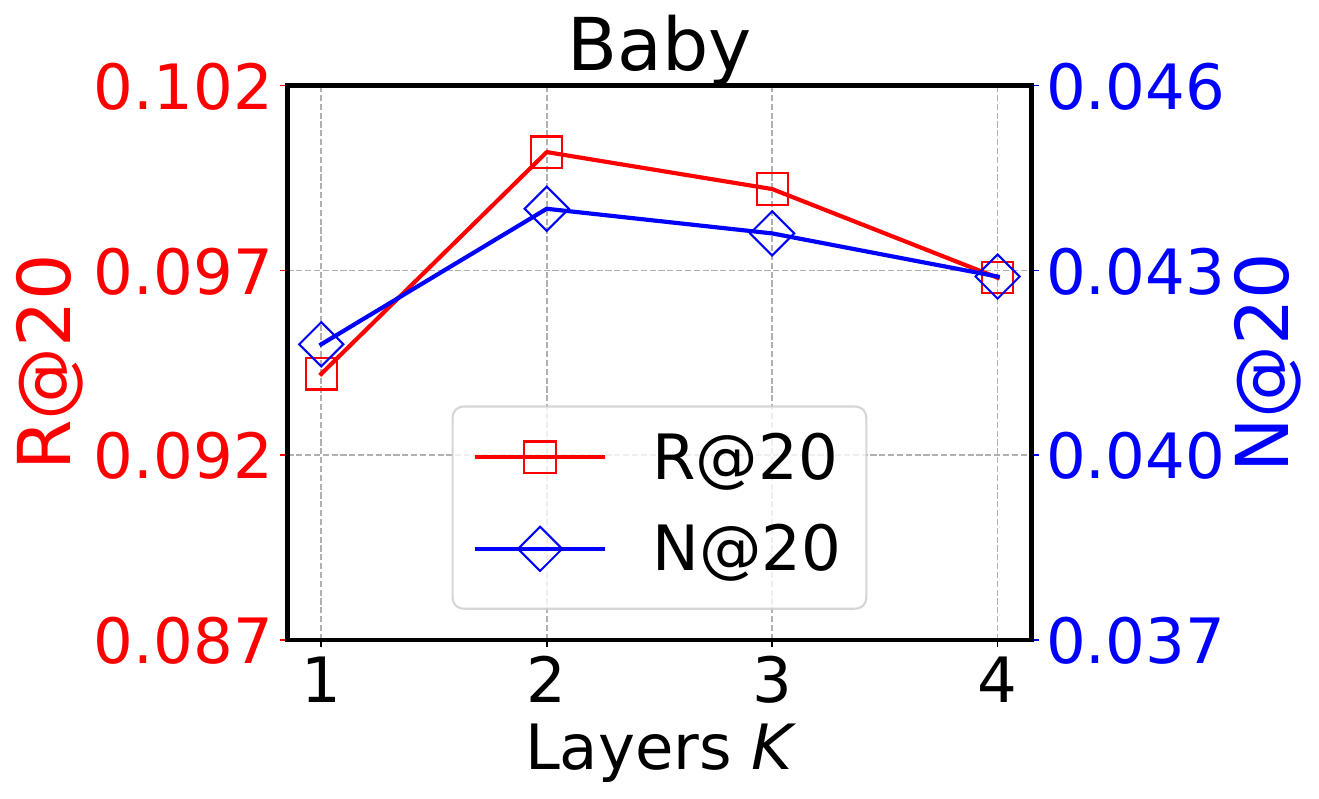}
		\includegraphics[width=0.16\linewidth]{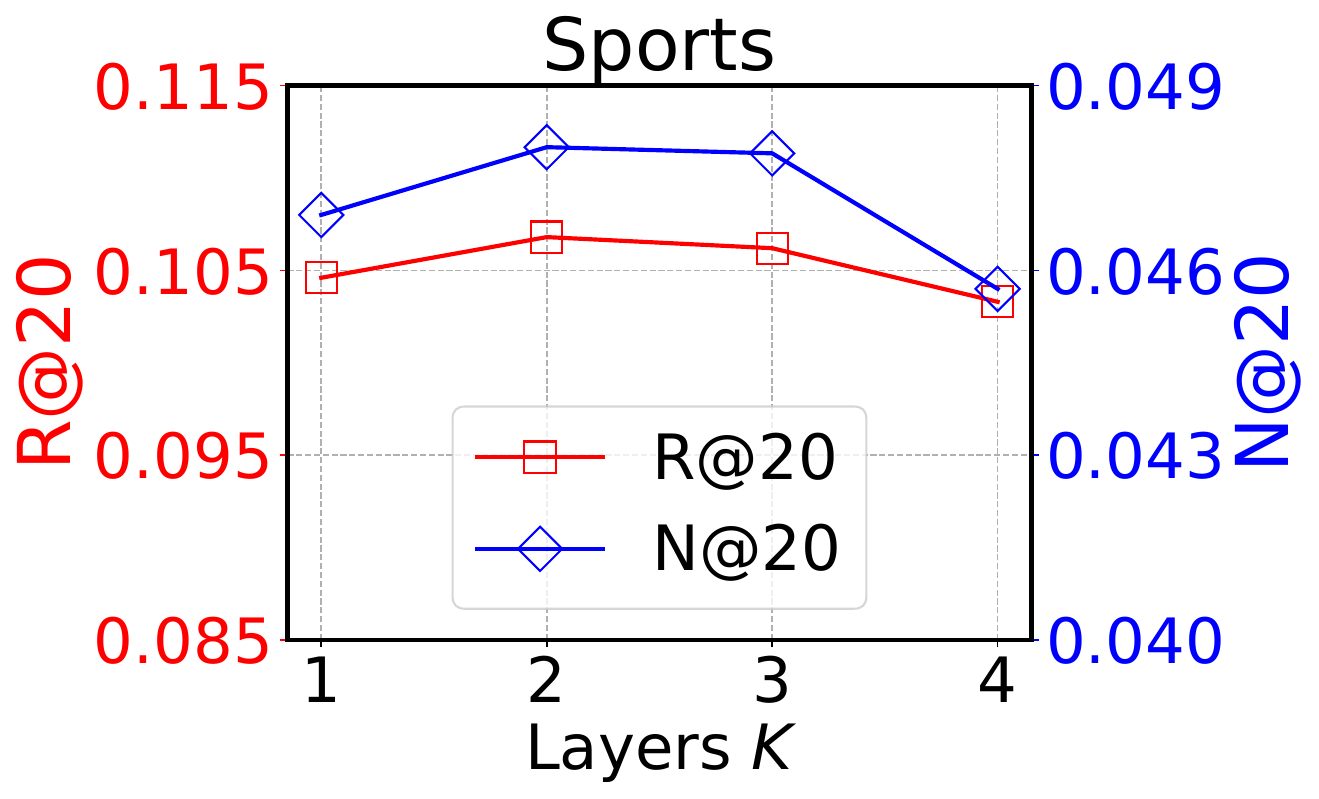}
		\includegraphics[width=0.16\linewidth]{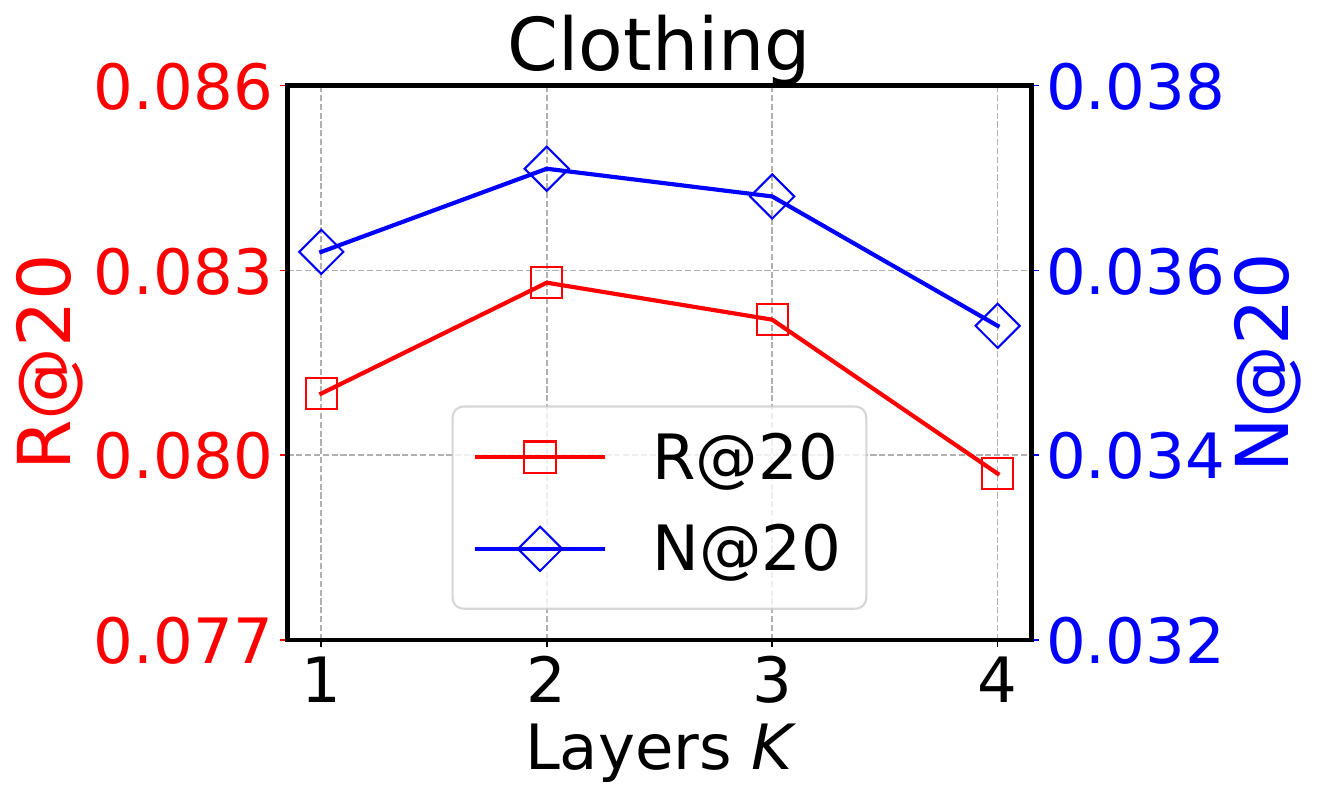}
    } \\
	\subfloat[\scriptsize{Impact of hypergraph layers $H$.}]{
		\includegraphics[width=0.16\linewidth]{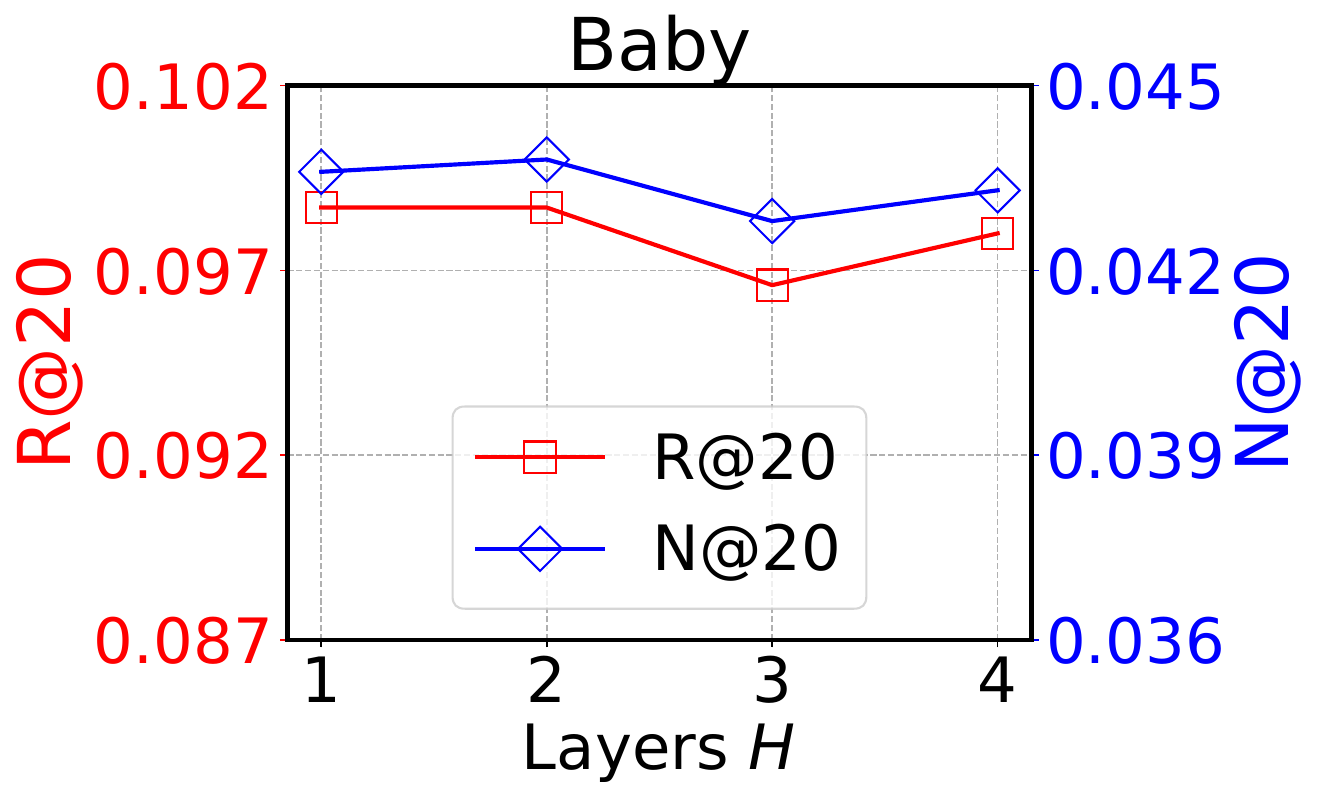}
		\includegraphics[width=0.16\linewidth]{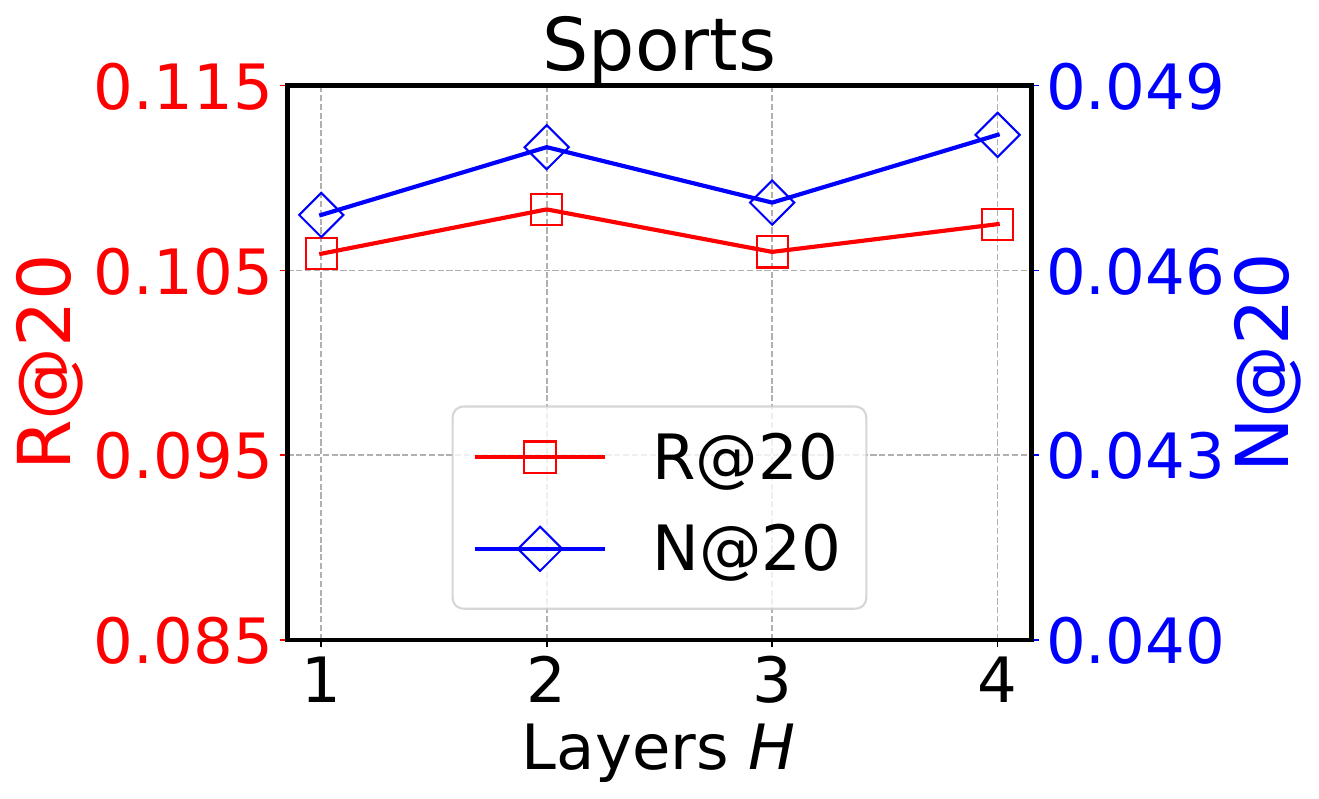}
		\includegraphics[width=0.16\linewidth]{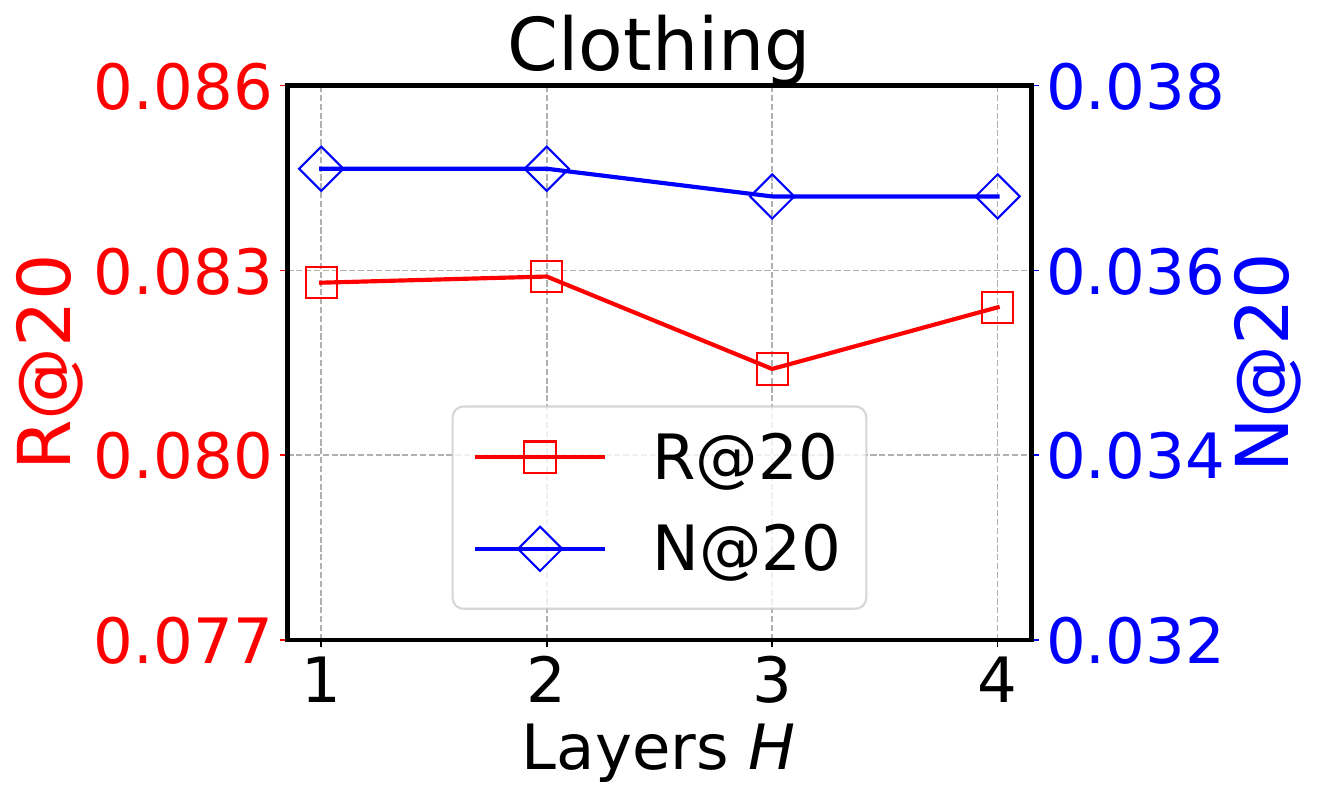}
	}
	\subfloat[\scriptsize{Impact of hyperedge number $A$.}]{
		\includegraphics[width=0.16\linewidth]{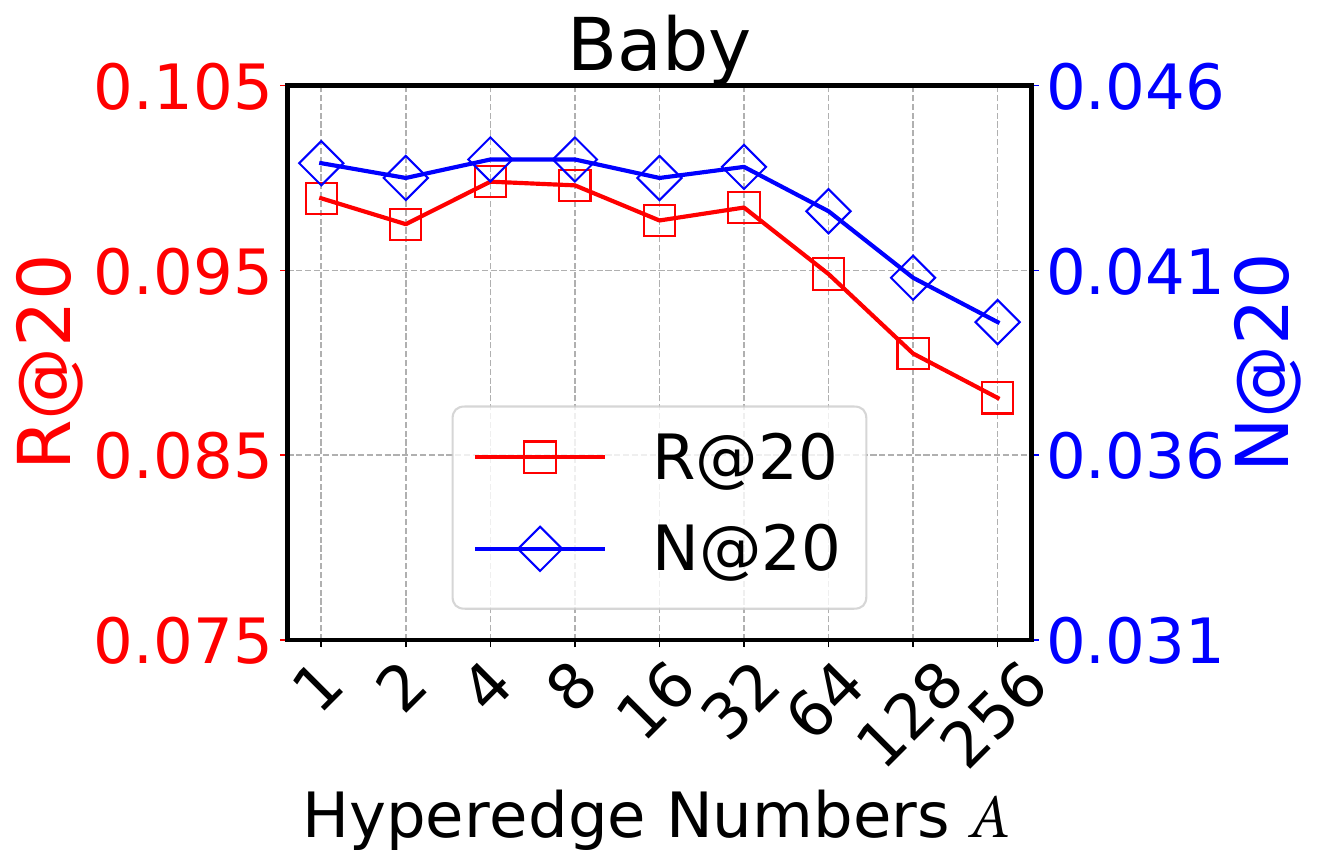}
		\includegraphics[width=0.16\linewidth]{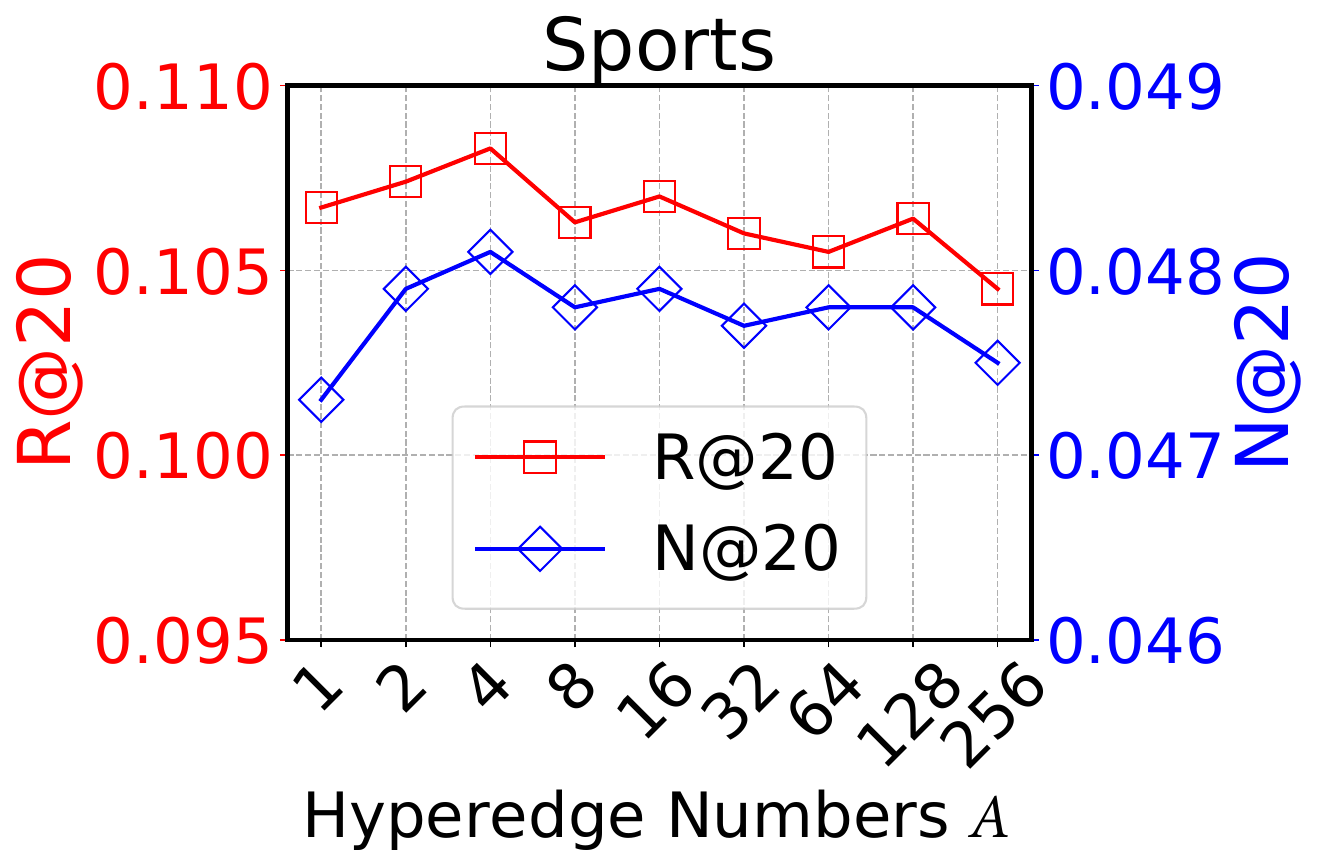}
		\includegraphics[width=0.16\linewidth]{image/Clothing-hyperedge.pdf}
	}\\
	\subfloat[\scriptsize{Impact of adjust factor $\alpha$.}]{
		\includegraphics[width=0.16\linewidth]{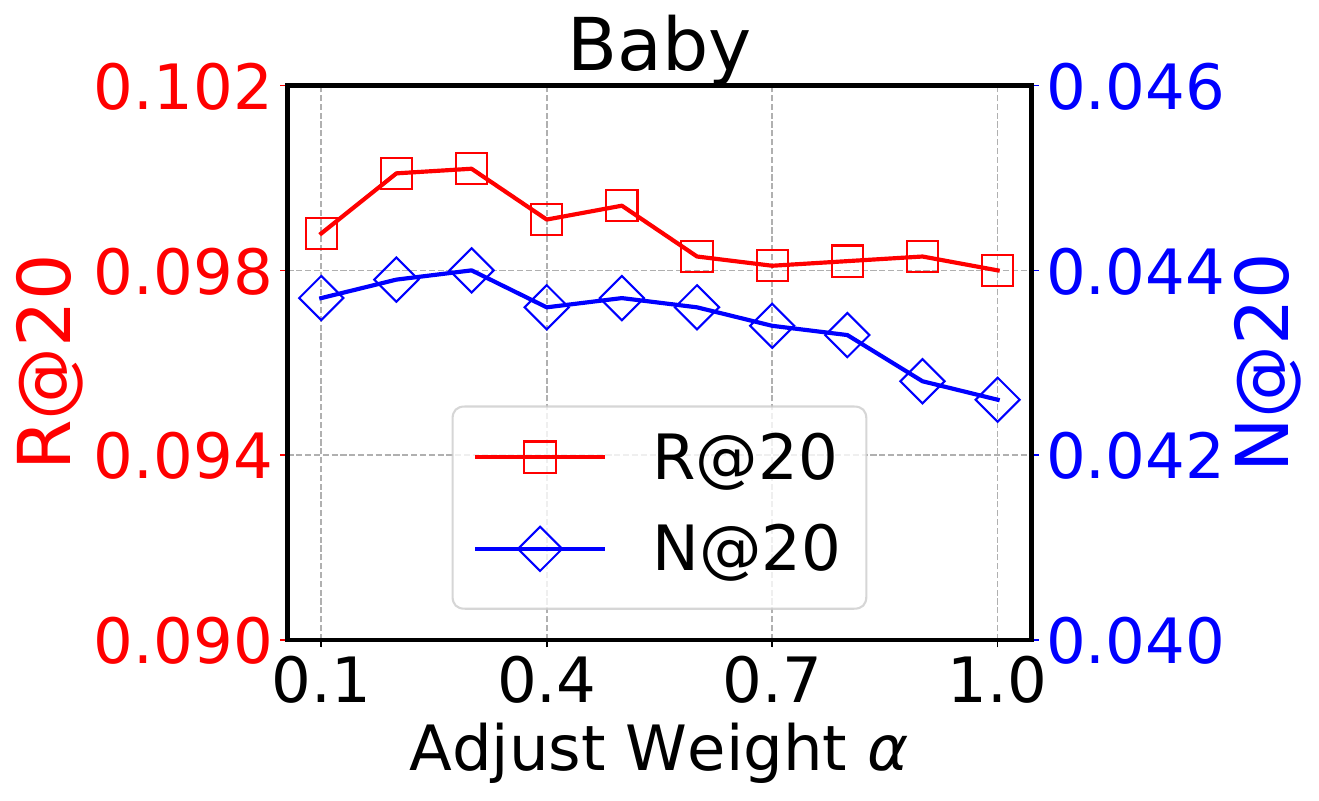}
		\includegraphics[width=0.16\linewidth]{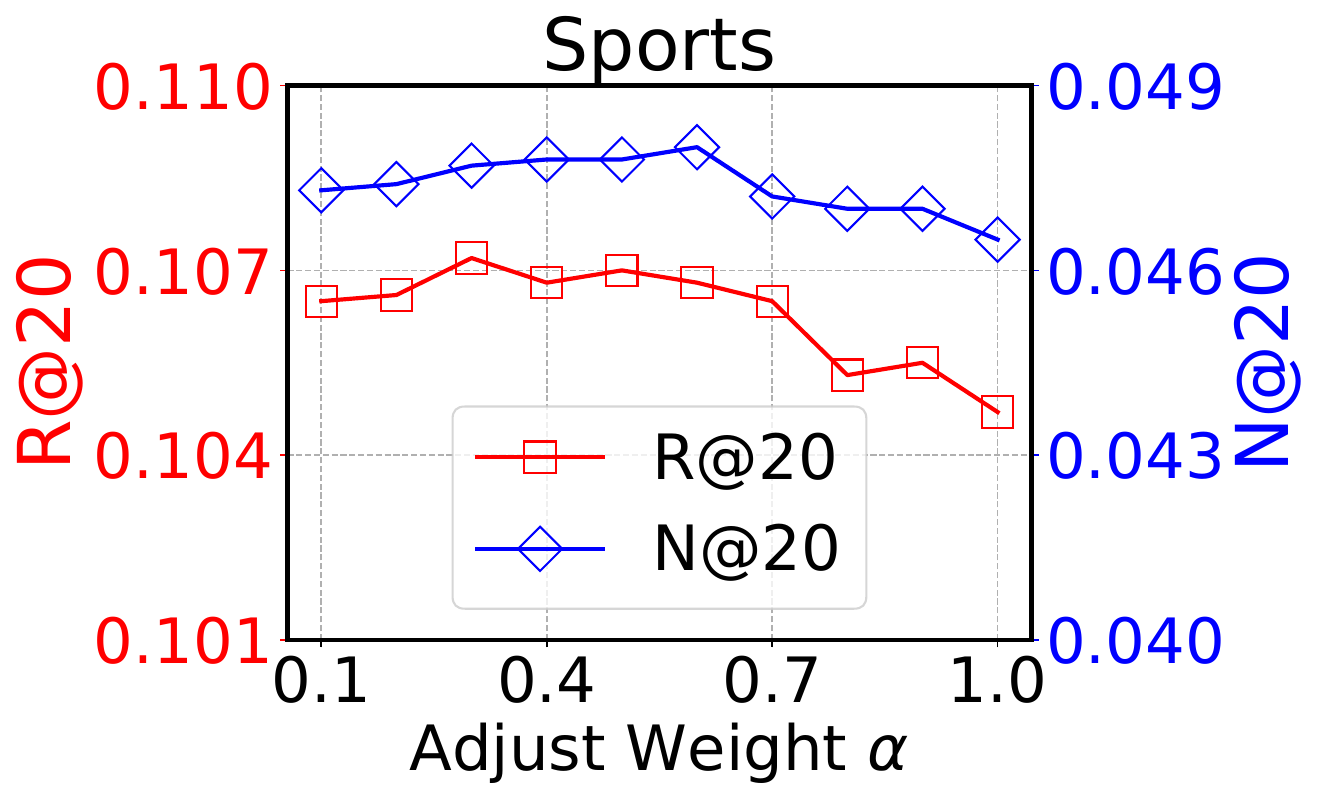}
		\includegraphics[width=0.16\linewidth]{image/Clothing-alpha.pdf}
	}
	\subfloat[\scriptsize{Impact of drop ratio $\rho$.}]{
		\includegraphics[width=0.16\linewidth]{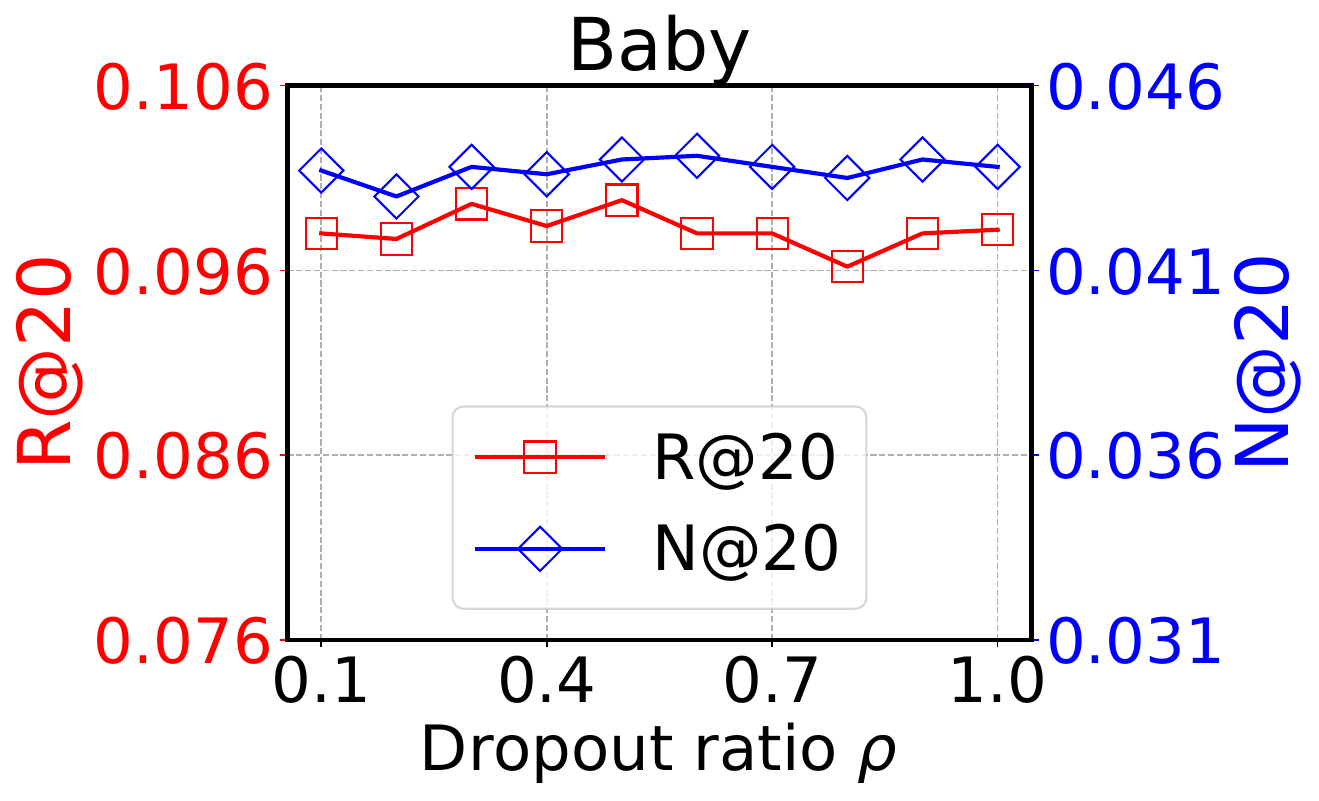}
		\includegraphics[width=0.16\linewidth]{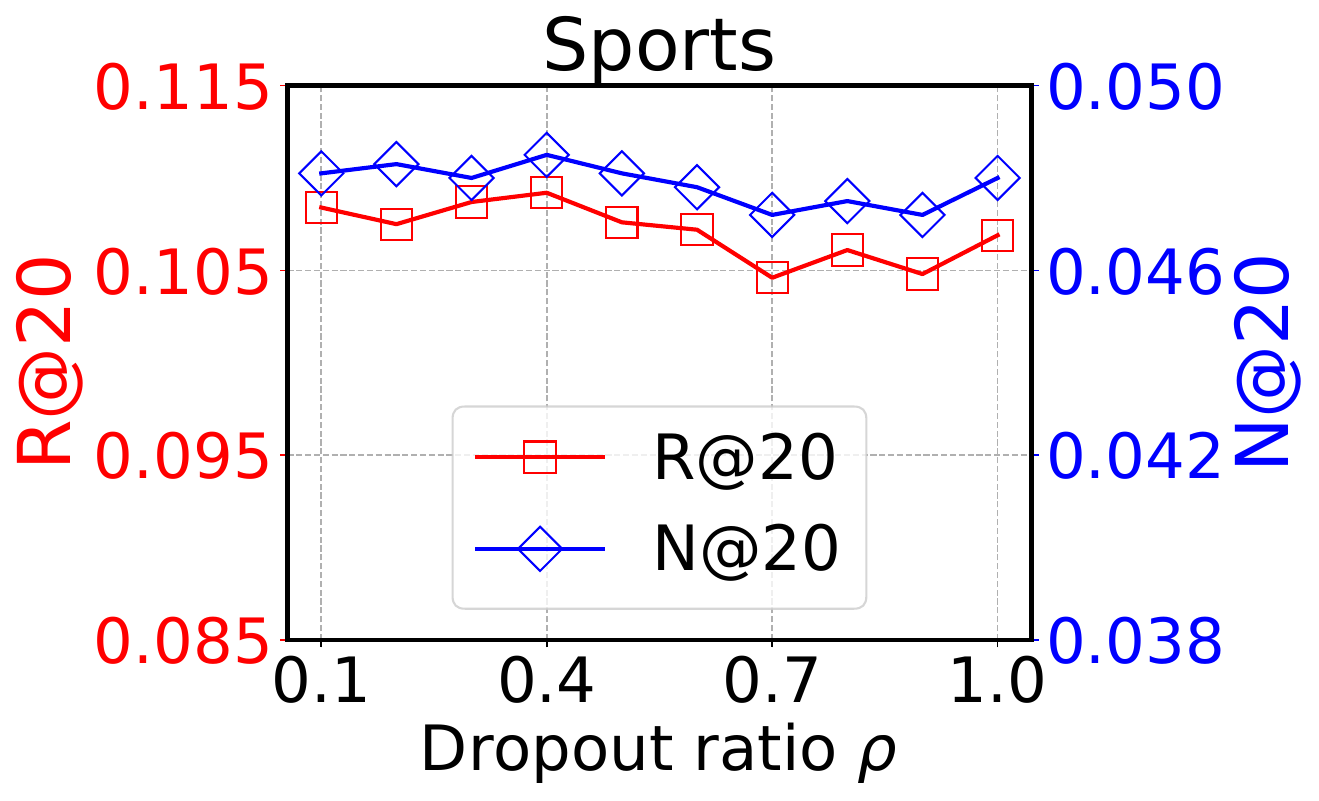}
		\includegraphics[width=0.16\linewidth]{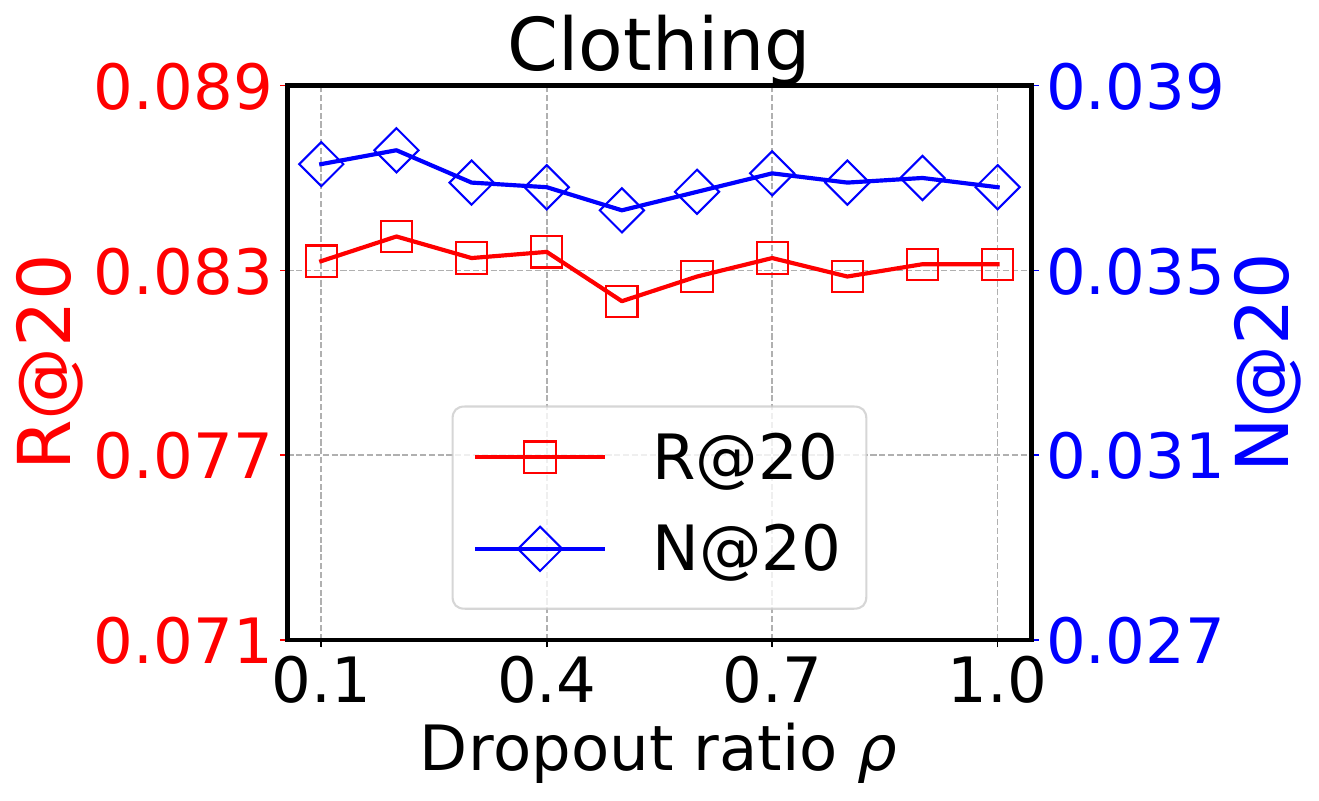}
	}\\
	\subfloat[\scriptsize{Impact of coefficient $\lambda_2$ of hypergraph contrastive learning.}]{
		\includegraphics[width=0.16\linewidth]{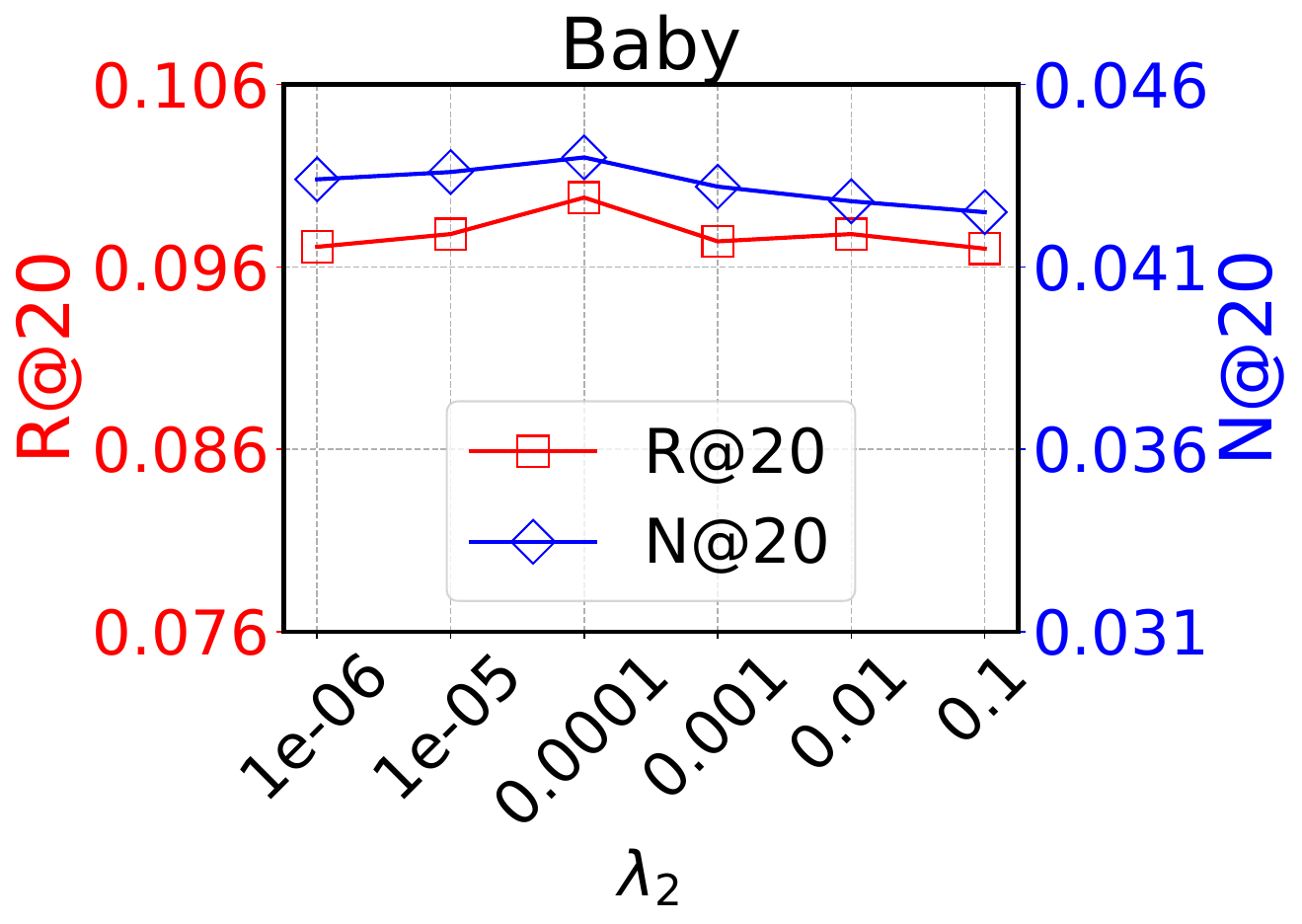}
		\includegraphics[width=0.16\linewidth]{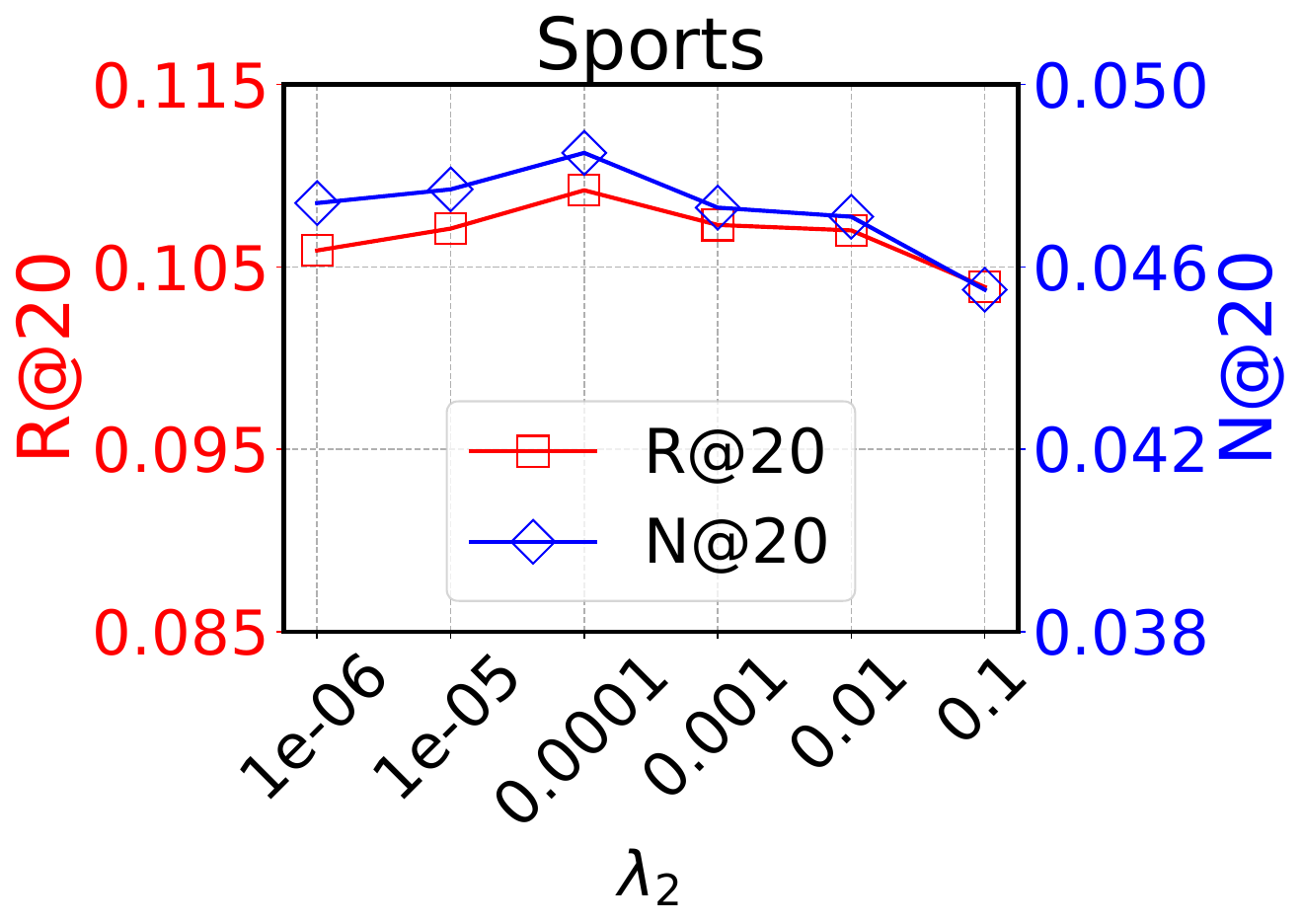}
		\includegraphics[width=0.16\linewidth]{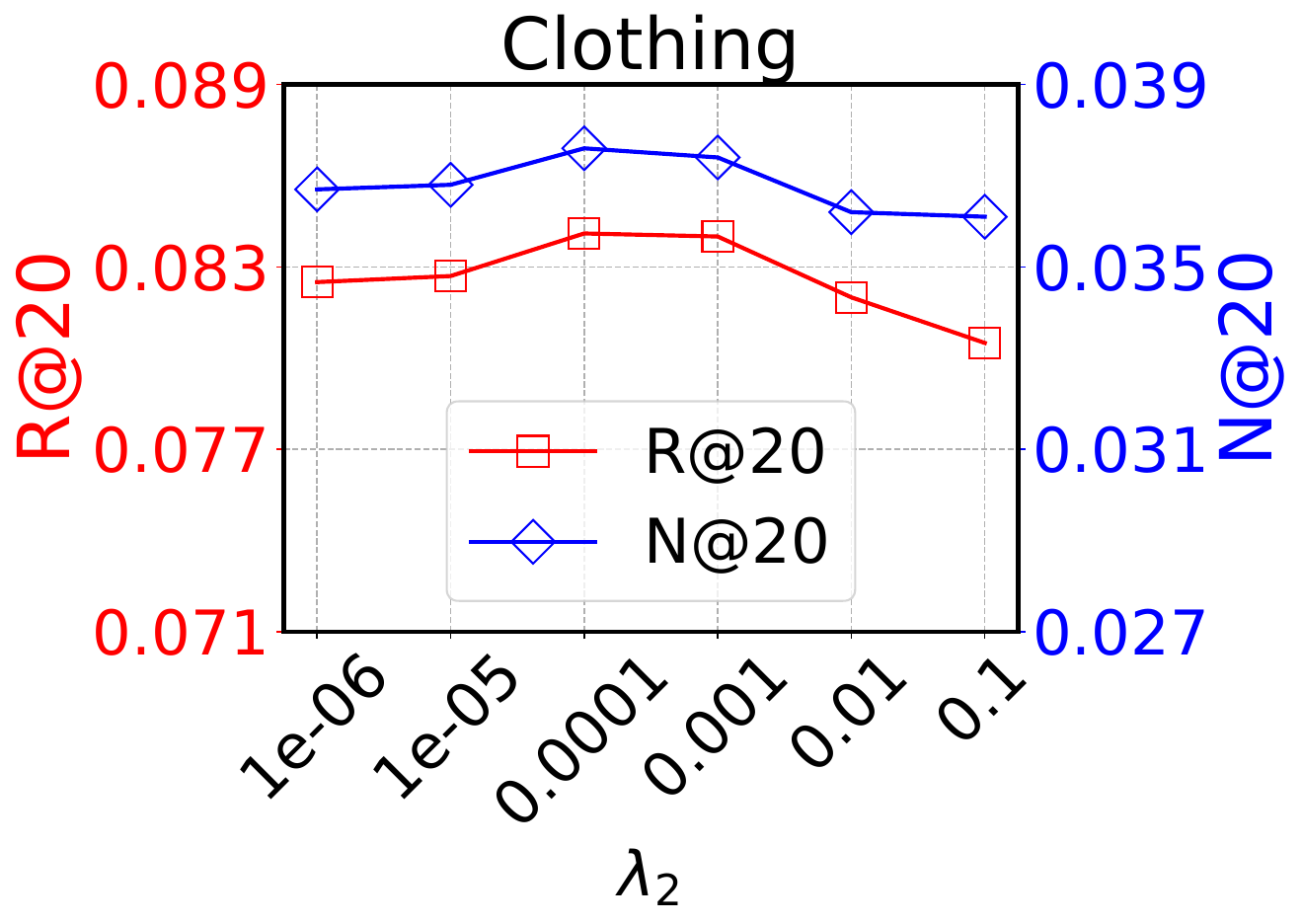}
	}
	\subfloat[\scriptsize{Impact of regularization coefficient $\lambda_1$.}]{
		\includegraphics[width=0.16\linewidth]{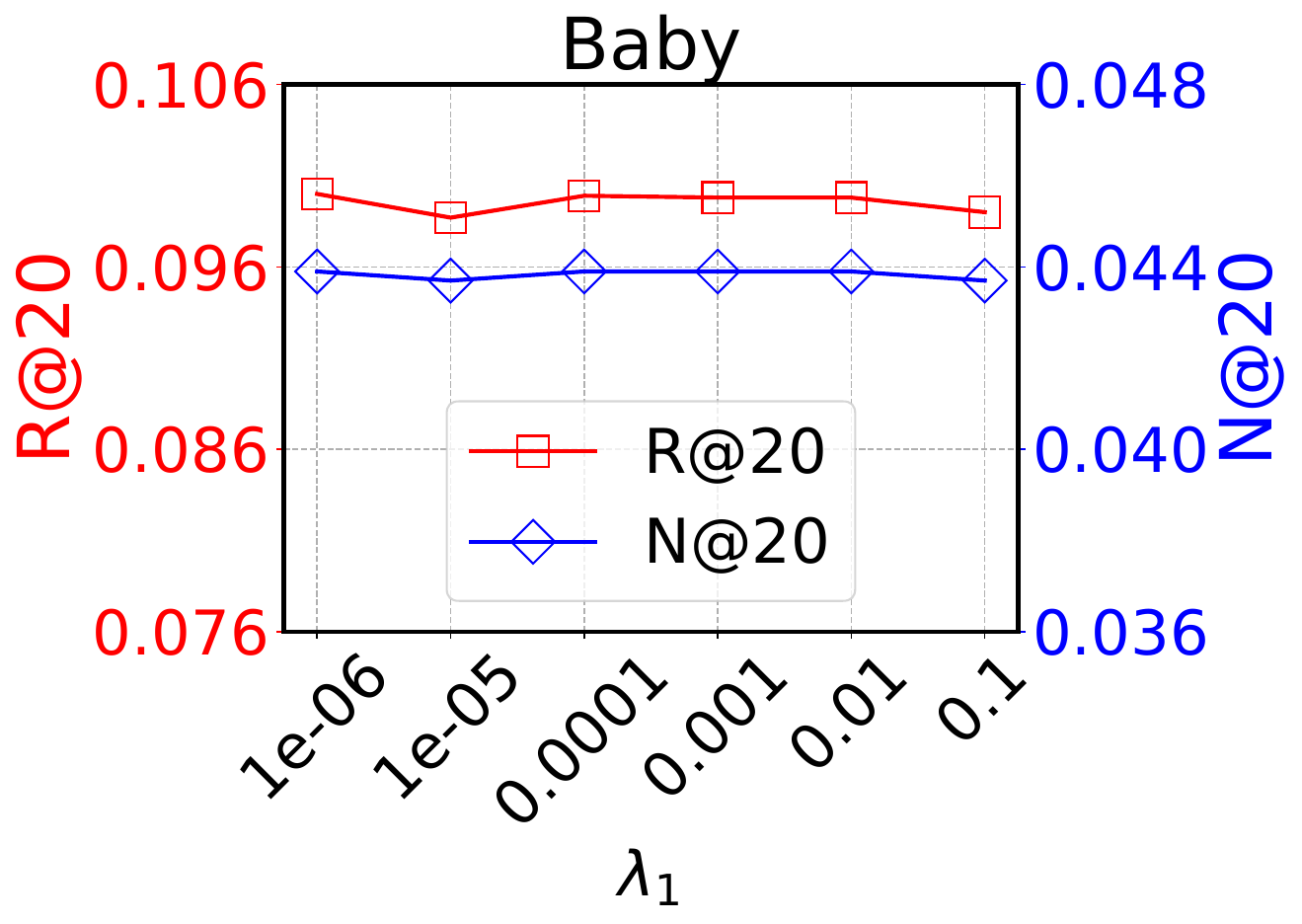}
		\includegraphics[width=0.16\linewidth]{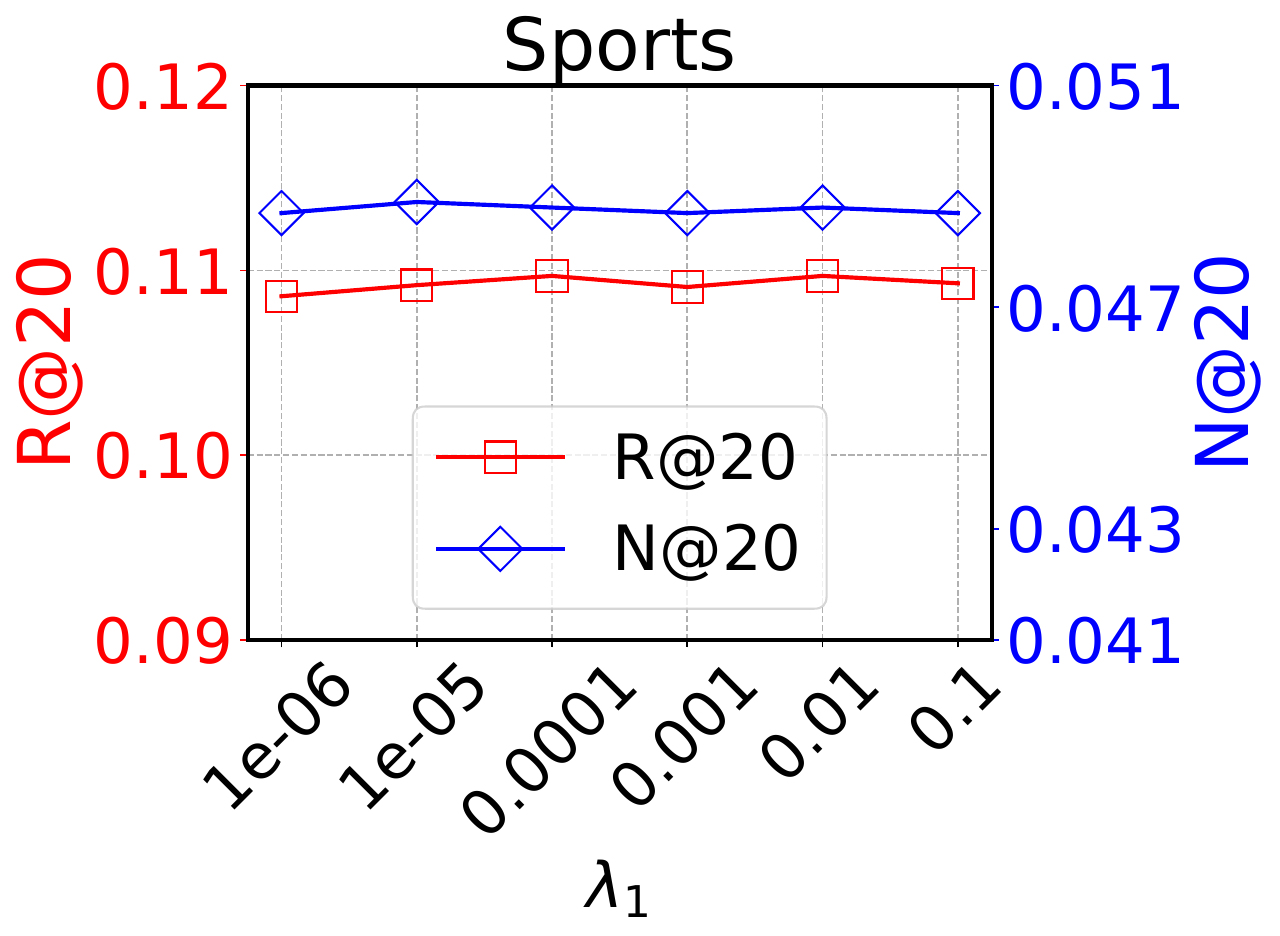}
		\includegraphics[width=0.16\linewidth]{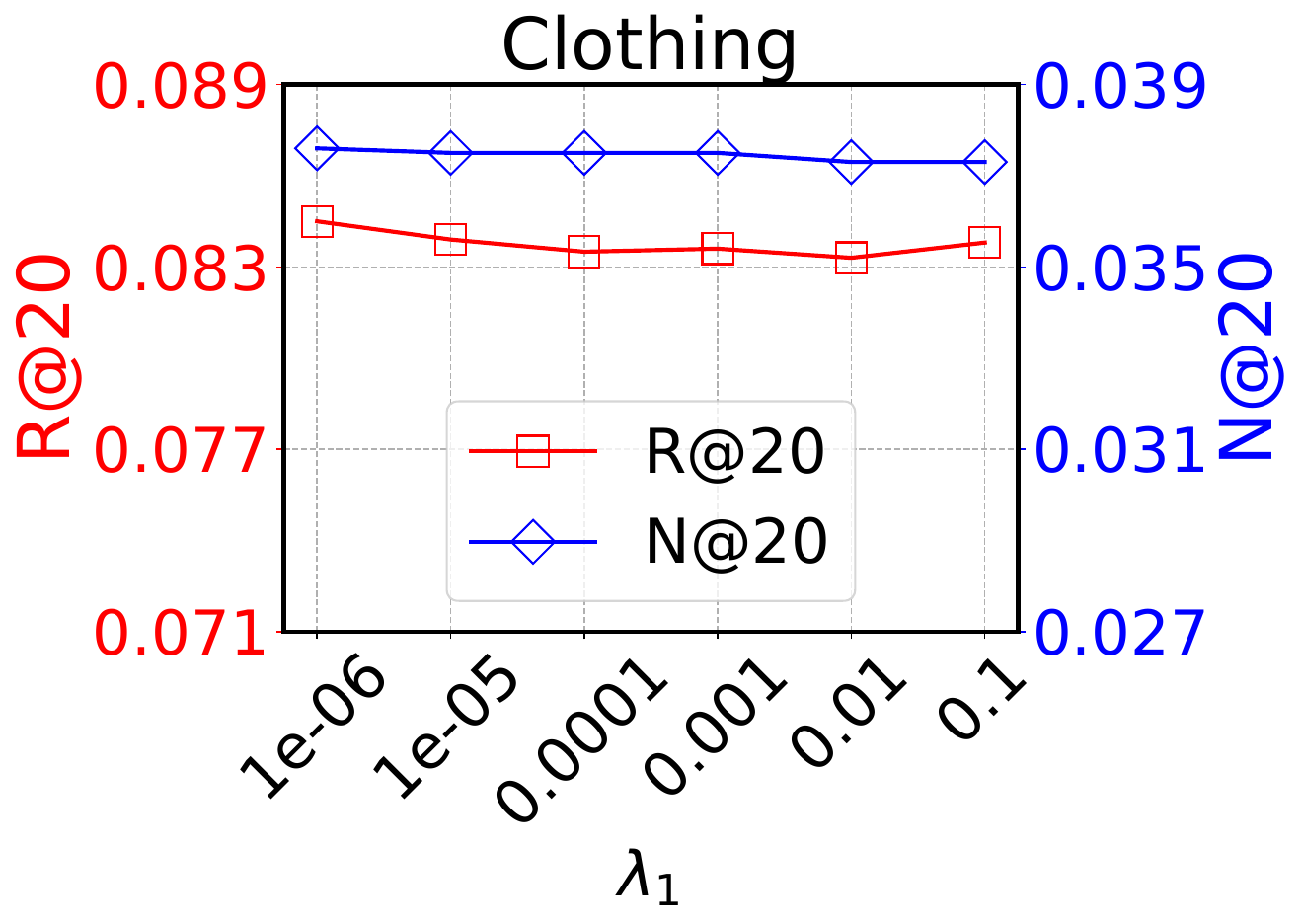}
	}\\
	\caption{Performances under different settings of all hyperparameters of LGMRec on Baby, Sports, and Clothing datasets.}
	\label{fig:all_parameters}
\end{figure*}

\subsection{Hyperparameter Analysis}
We explore the impacts of all hyperparameters for performance and report them in Figure~\ref{fig:all_parameters}.
\begin{itemize}[leftmargin=1em]
	\item \textbf{Collaborative graph layers $L$.} From Figure~\ref{fig:all_parameters}~(a), the results of the single-layer model are slightly inferior to that of the multi-layer model. The outcomes suggest that combining ID embeddings with sufficient multi-layer local structure information can obtain higher-quality user and item representations. In addition, the sparser datasets may require a deeper and larger local structure receptive field to facilitate recommendation, e.g., $L=4$ on Sports, and $L=3$ on Clothing.
	
	\item \textbf{Modality graph layers $K$.} The results in Figure~\ref{fig:all_parameters}~(b) demonstrate that the two-layer model achieves better performance and the increasing of layers does not bring a performance improvement, which indicates that the discrimination of nodes is decreasing as the layer number increases. The reason may be that aggregating deeper neighbors may lead to knowledge redundancy of node modality features.
	
	\item \textbf{Hypergraph layers $H$.} The impact of hypergraph layer $H$ is shown in Figure~\ref{fig:all_parameters}~(c). From the results, we can see that shallow global embedding performs better than multi-layer, possibly because multi-layer hypergraph propagation can lead to the excessive smoothness of node representations and reduce performance. In practice, we can take $H=1$ for all datasets.
	
	\item \textbf{Hyperedge number $A$.} Figure~\ref{fig:all_parameters}~(d) shows the performance of LGMRec with different settings of hyperedge number $A$. As mentioned in our experiments, the performance promotes as the number of hyperedges increases on the sparser Clothing dataset. For Baby and Sports datasets, the performance usually reaches its optimum at $A=4$. The results demonstrate the effectiveness of capturing multi-hyperedge global structures, especially for sparser datasets. 
	
	\item \textbf{Adjustable factor $\alpha$.} The performance of LGMRec with different settings of weight $\alpha$ is reported in Figure~\ref{fig:all_parameters}~(e). The results on the three datasets show a consistent trend, that is, the performance first increases to the optimal value and then decreases. The results suggest that properly supplementing global embeddings is suitable for modeling robust user interests. So, we can set $\alpha=0.3, 0.6, 0.2$ for Baby, Sports, and Clothing datasets, respectively.
	
	\item \textbf{Drop ratio $\rho$.} We tune the dropout ratio $\rho$ from $\{0.1,0.2, \ldots, 1.0\}$ to control the retention of hypergraph structure dependencies. The results in Figure~\ref{fig:all_parameters}~(f) indicate that proper dropout (compared to no dropout, i.e., $\rho=1.0$) is suitable to suppress the global noise and improve the robustness of the representation.
	
	\item \textbf{Coefficient $\lambda_2$.} Coefficient $\lambda_2$ determines the influence of hypergraph contrastive loss, and the performance of LGMRec under different  $\lambda_2$  is shown in Figure~\ref{fig:all_parameters}~(g). Similar to adjustable factor $\alpha$, the performance  also first improves  to reach optimal and then declines as $\lambda_2$ increases. The results illustrate that an appropriate $\lambda_2$ can mitigate value scale differences between HCL loss and BPR loss. In practice, we can uniformly set $\lambda_2=1e^{-4}$ on the three datasets.
	
	\item \textbf{Regularization coefficient $\lambda_1$.} We perform a grid search for parameter $\lambda_1$ to verify the effect of regularization. From Figure~\ref{fig:all_parameters}~(h), we can see that the effect of different coefficient $\lambda_1$ is negligible on the three datasets. Therefore, a small $\lambda_1=1e^{-6}$ is desirable.
\end{itemize}

\end{document}